\documentclass[12pt]{article}
\usepackage{epsfig}
\usepackage{amstex}
\usepackage{rotating}
\usepackage{dcolumn}
\usepackage{hhline}
\renewcommand{\baselinestretch}{1.1}
\newcommand{\al}{\alpha}

\newcommand{\ga}{\gamma}
\newcommand{\de}{\delta}
\newcommand{\ep}{\epsilon}
\newcommand{\varep}{\varepsilon}

\renewcommand{\th}{\theta}

\newcommand{\la}{\lambda}
\newcommand{\rh}{\rho}

\newcommand{\si}{\sigma}

\newcommand{\varph}{\varphi}
\newcommand{\ch}{\chi}
\newcommand{\ps}{\psi}
\newcommand{\om}{\omega}
\newcommand{\Ga}{\Gamma}

\newcommand{\klgl}{\:\hbox to -0.2pt{\lower2.5pt\hbox{$\sim$}\hss}
 {\raise3pt\hbox{$<$}}\:}
\newcommand{\grgl}{\:\hbox to -0.2pt{\lower2.5pt\hbox{$\sim$}\hss}
 {\raise3pt\hbox{$>$}}\:}

\def\bbbc{{\mathchoice {\setbox0=\hbox{$\displaystyle\rm C$}\hbox{\hbox
 to0pt{\kern0.4\wd0\vrule height0.9\ht0\hss}\box0}}
 {\setbox0=\hbox{$\textstyle\rm C$}\hbox{\hbox
 to0pt{\kern0.4\wd0\vrule height0.9\ht0\hss}\box0}}
 {\setbox0=\hbox{$\scriptstyle\rm C$}\hbox{\hbox
 to0pt{\kern0.4\wd0\vrule height0.9\ht0\hss}\box0}}
 {\setbox0=\hbox{$\scriptscriptstyle\rm C$}\hbox{\hbox
 to0pt{\kern0.4\wd0\vrule height0.9\ht0\hss}\box0}}}}
\newcommand{\fm}{{\rm \;fm}}
\newcommand{\GeV}{{\rm \;GeV}}
\newcommand{\milli}{\mbox{$\times\!10^{-3}$}}
\newcommand{\micro}{\mbox{$\times\!10^{-6}$}}
\newcommand{\non}{\mbox{---}}
\newcommand{\nn}{\nonumber}
\newcommand{\beq}{\begin{equation}}
\newcommand{\beqa}{\begin{eqnarray}}
\newcommand{\figeps}[2]{\epsfxsize=#1mm\epsfbox[72 240 540 540]{#2.eps}
  \setlength{\unitlength}{1mm}}
\newcommand{\figps}[2]{\epsfxsize=#1mm\epsfbox[100 300 530 570]{#2.ps}
  \setlength{\unitlength}{1mm}}
\begin{document}
\setcounter{page}{0}
\thispagestyle{empty}
%
%
%
\markboth{ }{ }
\renewcommand{\baselinestretch}{1.0}\normalsize
\hfill HD-THEP-98-24 \\
\vspace{\baselineskip}
\hfill HD-TVP-98-5 \\
\vspace*{1.5cm}
\renewcommand{\thefootnote}{\fnsymbol{footnote}}
\begin{center}
{\LARGE\bf Diffractive photo- and leptoproduction \\ \vspace*{0.1cm}
 of vector mesons \mbox{\boldmath $\rho$, $\rho'$} and \mbox{\boldmath$\rho''$}
\footnote{Supported by the Deutsche Forschungsgemeinschaft under grant no. GRK 216/1-96,
  by the EU grant FMRX-CT96-0008
  and by the Federal Ministry of Education, Science, Research and Technology
(BMBF), grant no. 06 HD 855} } \end{center}
\bigskip
\makeatletter \begin{center}
\large G.~Kulzinger\footnote{E-mail: G.Kulzinger@thphys.uni-heidelberg.de}, 
 H.G.~Dosch and H.J.~Pirner \vspace*{0.3cm} \\
{\it Institut f\"ur Theoretische Physik der Universit\"at Heidelberg \\
Philosophenweg 16 \& 19, D-69120 Heidelberg }
\end{center} \makeatother
{\begin{center} (June~12, 1998) \end{center}}
\vspace*{2cm}
\begin{abstract}
\noindent
We calculate diffractive photo- and leptoproduction of $\rh$-, $\rh'$- and $\rh''$-mesons.
The incoming photon dissociates into a $q\bar{q}$-dipole which scatters on the nucleon and
transforms into a vector meson state. The scattering amplitude is calculated in
non-perturbative QCD with the model of the stochastic vacuum. Assuming that the physical
$\rh'$- and $\rh''$-mesons are mixed states of an active 2S-excitation and some residual
hybrid state which cannot be produced diffractively in lowest order QCD, we obtain good
agreement with the data, especially the markedly different spectrum in the
$\pi^+\pi^-$-invariant mass for photoproduction and \mbox{$e^+e^-$-annihilation}.
\end{abstract}
\bigskip
{\begin{center}\it accepted for publication in Eur.Phys.J.{\bf C} \end{center}}

\newpage
\renewcommand{\baselinestretch}{1.1}\normalsize
\renewcommand{\thefigure}{\arabic{figure}}
\renewcommand{\thefootnote}{\arabic{footnote}}
\setcounter{footnote}{0}

\renewcommand{\thesection}{}
\section{Introduction} 
\setcounter{section}{0}

Exclusive vector meson production by real and virtual photons is an efficient
probe to investigate the physics of diffractive scattering. 
The experimental situation in $\pi^+\pi^-$- and $2\pi^+2\pi^-$-production in
the mass range from $1 \!-\! 2\GeV\/$ is rather complex. Photoproduction data 
show one broad bump in the $\pi^+\pi^-$-mass distribution \cite{Aston1} on the
upper tail of the $\rho$ at around $1.6 \GeV$.
The same enhancement is visible in $2\pi^+2\pi^-$-production \cite{Aston2}.
In $e^+e^-$-annihilation \cite{Q,Ba,Bi} a distinct interference pattern is seen.
Evidence for two resonances has been established in Refs~\cite{DM,CD}. 
Both resonances couple with approximately equal strength to the electromagnetic
current. Their masses are  compatible with those of the $1^{--}$ states
$\rh(1450)$ and $\rh(1700)$, respectively.

In Ref.~\cite{G1} good agreement with experimental data for $\rh$-production at
moderate and high photon virtualties $Q^2$ was obtained.
This success, based on the specific model of the stochastic vacuum for
non-perturbative QCD, sheds new light on the nature of the pomeron.
Since the stochastic gluon field strength correlators in the vacuum explain 
confinement, their application  to the physics of the pomeron builds an important
bridge between low-energy non-perturbative physics and high-energy scattering
at long distances. The coupling of the photon to the \mbox{$q\bar q$-dipole} 
is taken from perturbation theory. This approach has been checked in 
inclusive photon scattering at high $Q^2$ \cite{G2}, where at fixed
scattering energy $W$ the same photon wave function and reaction mechanism
reproduce the structure function $F_2$. 

For low $Q^2 \!<\! 1 \GeV^2$ the perturbative photon wave function is not acceptable,
since the resulting large $q \bar q$-dipoles feel confinement and chiral symmetry
breaking.
A way out of this dilemma has been shown in Ref.~\cite{G2}, where a $Q^2$-dependent
quark mass, determined from comparison with the phenomenological correlator of the
vector current, has been introduced in the perturbative photon wave function.
This effective mass mimics chiral symmetry breaking and also confinement in the
Euclidean region  as has been shown in a detailed model investigation of the harmonic
oscillator. 
Comparison with the phenomenological correlator indicates that chiral symmetry is
effectively restored at $Q^2 \!>\! 1 \GeV^2$, the constituent quark goes over into
a partonic massless quark.
Such a transition with resolution $Q^2$  is also seen in theoretical renormalization
flow equations \cite{JW, SchP}. It is intimately connected with the chiral phase
transition at finite temperatures. 
The calculation of diffractive vector meson production at low $Q^2$ in the following
paper will present an additional test of the validity for the chiral transition.

Vector dominance or generalized vector dominance could be in principle another
approach to treat the low-$Q^2$ virtual photon.
We found, however, that the method has little predictive power since the results
depend very strongly on couplings of the inserted vector meson states to the vector
current; also the number of inserted vector meson states influences the results
crucially.
This behaviour is not unexpected, because the construction of a transverse wave
function of a virtual photon, composed of wave functions of excited vector mesons,
has to imitate a delicate cancellation at large distances. 
A similar feature can be  seen very clearly in the harmonic oscillator model. 
In our opinion the modified perturbative $q\bar q$-wave function of the photon is
a more reliable and more predictive description of the low-$Q^2$ physics than the
treatment with generalized vector dominance.

%
%
\begin{figure}
$$
\figps{140}{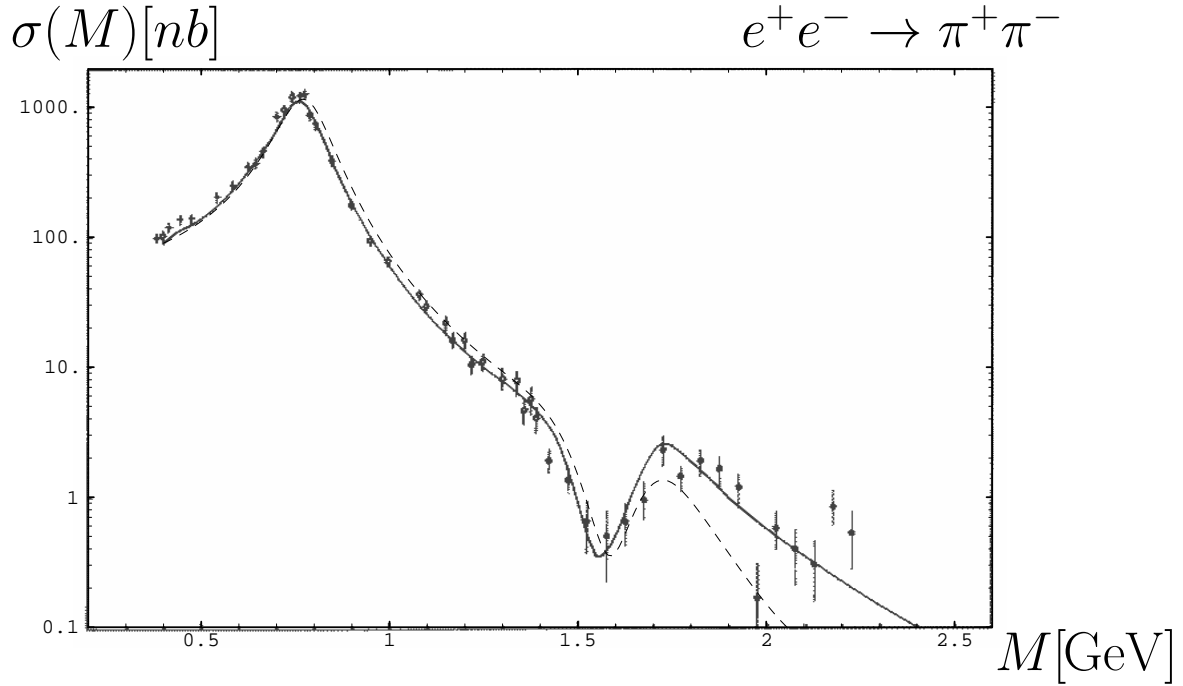}
\begin{picture}(0,0)
\put(3,6){\makebox(0,0){\LARGE $M[\!\!\GeV]$}}
\put(-140,89){\makebox(140,0){\LARGE $\si(M)[nb]$
    \hfill $e^+e^-\rightarrow\pi^+\pi^-$}}
\end{picture}
$$
\vspace*{-15mm}
\caption[]{ Mass spectrum of $e^+e^-$-annihilation into $\pi^+\pi^-$. In
  the 1.6\GeV\/ region a destructive interference shows up determining the
  sign pattern $(+,-,+)$ of the vector meson couplings $f_V$ to the
  electromagnetic current. The full curve is the fit of Donnachie
  and Mirzaie~\cite{DM}. The dashed line is the parametrization for $\rh'$
  and $\rh''$ used in this paper (see~Table~\ref{Tabl:properties} and
  App.~\ref{App:llbar}). } \label{Fig:eebar2pis}
\end{figure}
%

%
%
\begin{figure}
$$
\figps{140}{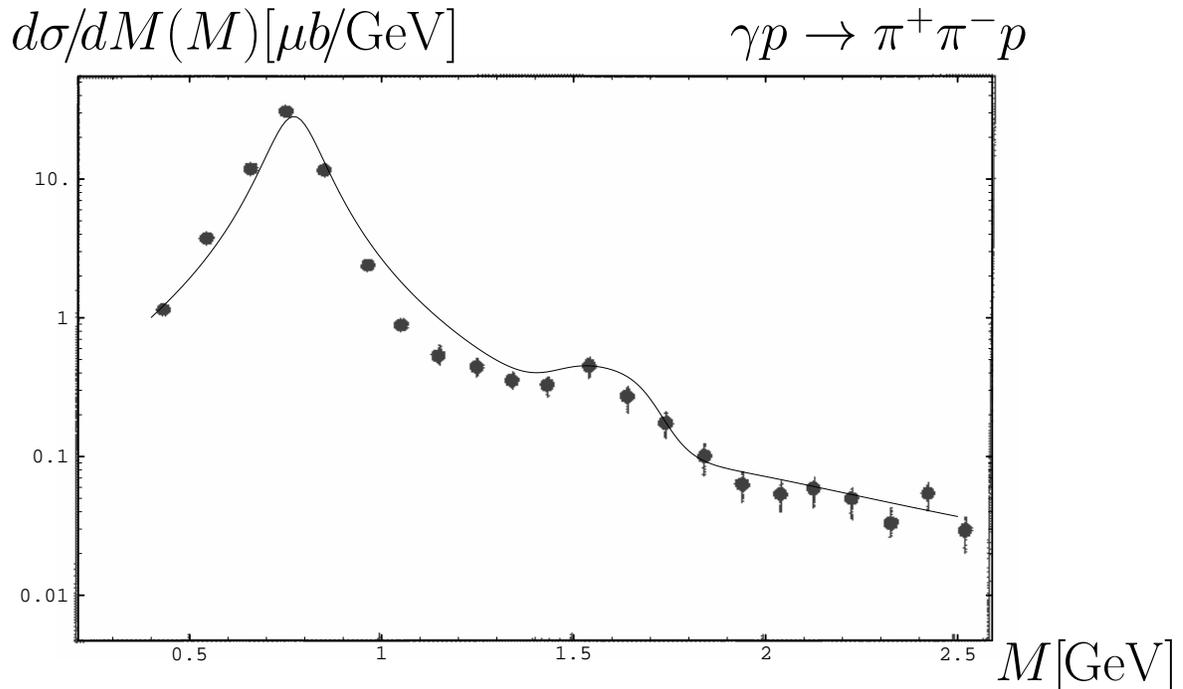}
\begin{picture}(0,0)
\put(3,6){\makebox(0,0){\LARGE $M[\!\!\GeV]$}}
\put(-140,90){\makebox(135,0){\LARGE $d\si\!/\!dM(M)[\mu b\!/\!\!\GeV]$
    \hfill $\ga p\rightarrow\pi^+\pi^-p$}}
\end{picture}
$$
\vspace*{-16mm}
\caption[]{ Mass spectrum of $\pi^+\pi^-$-photoproduction on the proton.
  The interference in the 1.6\GeV\/ region is constructive. The solid
  line is our result for $\pi^+\pi^-$-photoproduction using simple
  Breit-Wigner distributions for the $\rh$, $\rh'$ and $\rh''$. Experimental
  points are not normalized and taken from Aston et al.~\cite{Aston1}, with
  a contribution of $50\pm20$ nb from the $g(1690)$ subtracted~\cite{Atkinson}
  (see also~\cite{DM}). } \label{Fig:photo2pis}
\end{figure}

The outline of the paper is as follows: In Sect.~\ref{Sect:wfns} we give the
light-cone wave functions of the $\rh$, $\rh'$ and $\rh''$.
Sect.~\ref{Sect:wfns} also contains our comparison of theory with experimental
branching ratios and decay widths.
In Sect.~\ref{Sect:diffraction} we calculate the matrix elements and cross sections
for diffractive production of the vector meson states by real and virtual photons.
Sect.~\ref{Sect:discussion} concludes with a discussion and summary.

\renewcommand{\thesection}{\Roman{section}}
\section{\hspace*{-.85em}. Wave functions and properties of
  \mbox{\boldmath $\rho$, $\rho'$} and \mbox{\boldmath$\rho''$}} \label{Sect:wfns}

\renewcommand{\thesubsection}{\alph{subsection}.}
\subsection{Light-cone wave functions}

The hadronic light-cone wave functions of the mesons represent an important
input to exclusive scattering.
In the perturbative regime for longitudinal photons at high $Q^2$ the wave
function at the origin dominates the production process. This value of the wave
function is known from the measured $e^+e^-$-width. Parametrizations of the ground
state vector mesons based on this empirical information have been developed
in Ref.~\cite{G1}.
Here we want to extend this work to the excited light vector mesons $\rh'$ and
$\rh''$.
An analysis of the experimental data from $e^+e^-$-annihilation and photoproduction
of $(\pi^+,\pi^-)$ shows that there are at least two excited $\rh$-resonances, the
$\rh(1450)$ and the $\rh(1700)$~\cite{DM}.
Recently \cite{CD,Close} it has been speculated that there may be a hybrid state
$h(1450)$ with the quantum numbers of the $\rh$-meson which decays predominantly
into $\pi a_1$. 

The genuine quark model states are the $2S$- and $2D$-excitations. The $2S$-state
couples to the photon strongly, whereas the $2D$-state has a vanishing wave function
at the origin and consequently only a small relativistically induced coupling to the
photon \cite{Be}.
Also diffraction proceeds mostly without angular momentum transfer, so the production
of the $2D$-state is suppressed. 
In the following we will use a simplified ansatz for the vector meson states.
We employ the nonrelativistic notation $1S$ and $2S$ as a short hand notation for
light-cone wave functions which in the nonrelativistic limit have this character.
Our ansatz for the physical vector meson states has the following form:
\beqa \label{mixing}
|\rho(770)\rangle\;\, &=& \;\;\; |1S\rangle \;, \nn \\
|\rho(1450)\rangle    &=& \;\;\; \cos\theta \; |2S \rangle
                               + \sin\theta \; |rest \rangle \;, \nn \\
|\rho(1700)\rangle    &=&       -\sin\theta \; |2S \rangle
                               + \cos\theta \; |rest \rangle \;.
\end{eqnarray}
Here the state $|rest\rangle$ describes the $|2D\rangle$- and hybrid
$|h\rangle$-states whose coupling to the photon are suppressed and which hence
we neglect in our approach. For details of the wave functions both for the photon
and the vector mesons we refer to App.~\ref{App:wave_fns}.

\paragraph{Photon wave function.} For the photon wave function we use the
form derived in Ref.~\cite{G1} with a running quark mass $m(Q^2)$ in order
to take into account chiral symmetry breaking and confinement at large
distances in an approximate way. It depends on the light-cone momentum
fraction $z$ of the quark and the transverse distance $r$ between the quark
and the antiquark. The  index $\la$ indicates the helicity of the photon,
$h$ and $\bar h$ give the quark and antiquark helicities: 
\beq \label{photon1}
\ps_{\ga(Q^2,\la)}(z,{\bf r}) = 
   \sqrt{N_c}\; e_f\,\delta_{f\bar f}\; \ch_{\ga(Q^2,\la)}(z,{\bf r}) \;,
\end{equation}
with\footnote{In the following some indices and arguments are not given explicitely
in order not to overload the notation.}
\beqa \label{photon2}
\ch_{\ga(Q^2,\la=0)} &=& -\, \delta_{h,-\bar h}\; 2z(1-z)\; Q\;
   \frac{K_0(\varep r)}{2 \pi} \;, \\
\ch_{\ga(Q^2,\la=+1)} &=& \sqrt{2}\,
   \bigg\{\; i {\rm e}^{i\varph}\; 
   \varep \left( z \delta_{h+,\bar h-} - (1-z) \delta_{h-,\bar h+} \right)
   \frac{K_1(\varep r)}{2 \pi}
  + m(Q^2)\, \delta_{h+,\bar h+}\,
   \frac{K_0(\varep r)}{2 \pi} \bigg\} \;, \nn \\
\ch_{\ga(Q^2,\la=-1)} &=& \sqrt{2}\,
   \bigg\{ i {\rm e}^{-i\varph}
   \varep \left( (1-z) \delta_{h+,\bar h-} - z \delta_{h-,\bar h+} \right)
   \frac{K_1(\varep r)}{2 \pi}
  + m(Q^2)\, \delta_{h-,\bar h-}\,
   \frac{K_0(\varep r)}{2 \pi} \bigg\} \;, \nn 
\end{eqnarray}
where $\varph$ is the azimut angle and
\beq \label{epsilon}
\varep = \sqrt{z(1-z) Q^2 +m^2(Q^2)} \;.
\end{equation}
The running quark mass was determined to evolve as
\beq \label{mQ2}
m(Q^2) \;=\; \left\{ 
  \begin{array}{rl} 0.220 \GeV \cdot \left(1 - Q^2 /Q_0^2 \right),
       &\qquad Q^2 < Q_0^2 = 1.05 \GeV^2 \;, \\
    0, &\qquad Q^2 \ge Q_0^2 \;,
  \end{array} \right.
\end{equation}
in Ref.~\cite{G2} by matching the vector current correlator. 

\paragraph{Vector meson wave functions.} The vector meson light-cone wave
functions are parametrized in an analogous way. One has to rely on such a
phenomenological construction as long as 
not even the form of the light-cone Hamiltonian for valence states is known.
However, there are attempts to construct light-cone wave functions
via a Melosh transformation from solutions of a relativized constituent quark 
model Hamiltonian \cite{Simula}.
Recently, also a string equation for the meson on the light-cone
has been solved \cite{MoShSi}.
Since both approaches have not specified the solutions for the vector states
in a parametrized form, we use model wave functions similar to those of
Wirbel and Stech~\cite{WS} to set up the wave functions for the quark-antiquark
$1S$- and $2S$-excitations. Since the contributions of $z$ near the
endpoints are not significant for production of vector mesons at moderate $Q^2$
the argument against factorization of \cite{HaZh} and \cite{BaBr} have here no
practical consequence, see also \cite{Di}.

For convenience we introduce the following abbreviations:
\beqa \label{hg_defs}
h_{V, \la}(z)  &=& {\cal N}_{V,\la} \; \sqrt{z(1-z)} \;
  \exp\left\{ -\frac{1}{2}\,
    \frac{M^2 (z-1/2)^2}{\om_{V,\la}^2} \right\} \;, \\
g_{V, \lambda}(r) &=& \exp\left\{ -\frac{1}{2}\, \om_{V,\la}^2\, r^2 \right\} \;,
\end{eqnarray}
where $\la \!=\! L,T$ refers to longitudinal and transverse polarization and
$V \!=\!1, 2$ to the $1S$- and $2S$-state; $M$ is the mass of the $1S$-state, i.e.
of the $\rh$-meson. We have the following wave functions:

\noindent
\mbox{\boldmath$1S$}, {\bf longitudinal}:
\beq \label{wfnRhL}
\psi_{1(L)} = \delta_{h,-\bar h}\; 4z(1-z)\; \omega_{1L}\; h_{1L}(z)\, g_{1L}(r) \;.
\end{equation}
\mbox{\boldmath$1S$}, {\bf transverse}:
\beqa \label{wfnRhT}
\psi_{1(\la=+1)} &=& h_{1T}(z) \, g_{1T}(r) \\
   && \times \left\{\; i \omega_{1T}^2 r {\rm e}^{i\varph}\;
   \left( z \delta_{h,+}\, \delta_{\bar h,-}
       - (1- z) \delta_{h,-} \, \delta_{\bar h,+} \right)
   + m(Q^2)\,\delta_{h,+} \, \delta_{\bar h,+} \right\} \;, \nn \\
\psi_{1(\la=-1)} &=& h_{1T}(z) \, g_{1T}(r) \nn \\
   && \times \left\{ i \omega_{1T}^2 r {\rm e}^{-i\varph}
   \left( (1-z) \delta_{h,+} \, \delta_{\bar h,-} 
         - z \delta_{h,-} \, \delta_{\bar h,+} \right)
   + m(Q^2)\, \delta_{h,-} \, \delta_{\bar h,-} \right\} \;. \nn
\end{eqnarray}
For the $2S$-state we allow for an excitation in the transverse plane by taking
in momentum space the excited two dimensional harmonic oscillator wave function,
for the excitation in the 3-direction we introduce a polynomial quadratic in
$z$ and symmetric under interchange of $z$ and $(1\!-\!z)$. It is further fixed
by the condition that the $2S$-state is orthogonal on the $1S$-state.
We thus obtain:

\noindent
\mbox{\boldmath$2S$}, {\bf longitudinal}:
\beq \label{wfn2SL}
\hspace*{-1em}
\psi_{2(L)} = \delta_{h,-\bar h}\; 4 z(1-z)\; \omega_{2L}\; h_{2L}(z)\, g_{2L}(r)\,
  \left\{ (z(1-z) - A_L) + \sqrt{2}(\omega_{2L}^2 r^2 -1) \right\} \;.
\end{equation}
\mbox{\boldmath$2S$}, {\bf transverse}:
\beqa \label{wfn2ST}
\hspace*{-.5em}
\psi_{2(\la=+1)} &\hspace*{-.5em}=\hspace*{-.5em}& h_{2T}(z) \, g_{2T}(r) \\
   &&\times \bigg\{\; i \omega_{2T}^2 r {\rm e}^{i\varph}\;
   \left( z \delta_{h,+} \, \delta_{\bar h,-}
       - (1-z) \delta_{h,-} \, \delta_{\bar h,+} \right)
   \left[ (z(1-z) - A_T) + \sqrt{2}\, (\omega_{2T}^2 r^2 - 3) \right] \nn \\
   &&\hspace{11.125em} +\, m(Q^2)\, \delta_{h,+} \, \delta_{\bar h,+}
   \left[ (z(1-z) - A_T) + \sqrt{2}\, (\omega_{2T}^2 r^2 - 1) \right] \bigg\} \;, \nn \\
\psi_{2(\la=-1)} &\hspace*{-.5em}=\hspace*{-.5em}& h_{2T}(z) \, g_{2T}(r) \nn \\
   &&\times \bigg\{ i \omega_{2T}^2 r {\rm e}^{-i\varph}
   \left( (1-z) \delta_{h,+}\, \delta_{\bar h,-}
         - z \delta_{h,-}\, \delta_{\bar h,+} \right)
   \left[ (z(1-z) - A_T) + \sqrt{2}\, (\omega_{2T}^2 r^2 - 3) \right] \nn \\
   &&\hspace{11.125em} +\, m(Q^2)\, \delta_{h,-} \, \delta_{\bar h,-}
   \left[ (z(1-z) - A_T) + \sqrt{2}\, (\omega_{2T}^2 r^2-1) \right] \bigg\} \;; \nn
\end{eqnarray}
the factor $\sqrt{2}$ accounts for the two transverse excitation modes.

The normalization constants ${\cal N}_{1\la}$  are fixed by the wave function
normalization. The oscillator  frequencies $\omega_{1\lambda}$ are chosen in
such a way as  to reproduce the $\rh$-meson electromagnetic decay coupling
\mbox{$f_{1L}=f_{1T}$}. The values for $\omega_{2\lambda}$ were minimally 
deviated from the $1S$-values in order to give the same $2S$-leptonic coupling
for the longitudinal and transverse state. The constants ${\cal N}_{2\la}$ and
$A_L$, $A_T$ are  determined by the requirement that the $2S$-state is both
normalized and orthogonal on the $1S$-state. For details we refer to
App.~\ref{App:wave_fns}. In Table~\ref{Tabl:wf_cnsts} we collect the relevant
parameters. 

\newcolumntype{f}[1]{D{.}{.}{#1}}
\begin{table} 
\begin{center}
\begin{tabular}{|l||c|c|c|f{3}|}
\hhline{|-||----|}
\multicolumn{1}{|l||}{state}
  & \multicolumn{1}{c|}{$f_{V,\la} [\!\!\GeV]$}
  & \multicolumn{1}{c|}{$\om_{V,\la}[\!\!\GeV]$}
  & \multicolumn{1}{c|}{${\cal N}_{V,\la}$}
  & \multicolumn{1}{c|}{$A_\la$} \\
\hhline{:=::====:}
$1S$-Longitudinal & {\bf 0.1526} &      0.330  & 4.48 & \\
$1S$-Transverse   & {\bf 0.1526} &      0.213  & 3.44 & \\
$2S$-Longitudinal &  --\,0.137   & {\bf 0.297} & 3.21 &    0.228 \\
$2S$-Transverse   &  --\,0.137   & {\bf 0.235} & 1.96 & -\,0.328 \\
\hhline{|-||----|}
\end{tabular}
\end{center}
\caption{The parameters for the $1S$- and $2S$-wave functions.
  Besides the $\rh$-mass the bold face quantities are input.
  The values $\om_{2\la}$ are adjusted in order to have agreement
  of $f_{2L}$ with $f_{2T}$} \label{Tabl:wf_cnsts}
\end{table}

%
%
\begin{figure}
$$
\figeps{138}{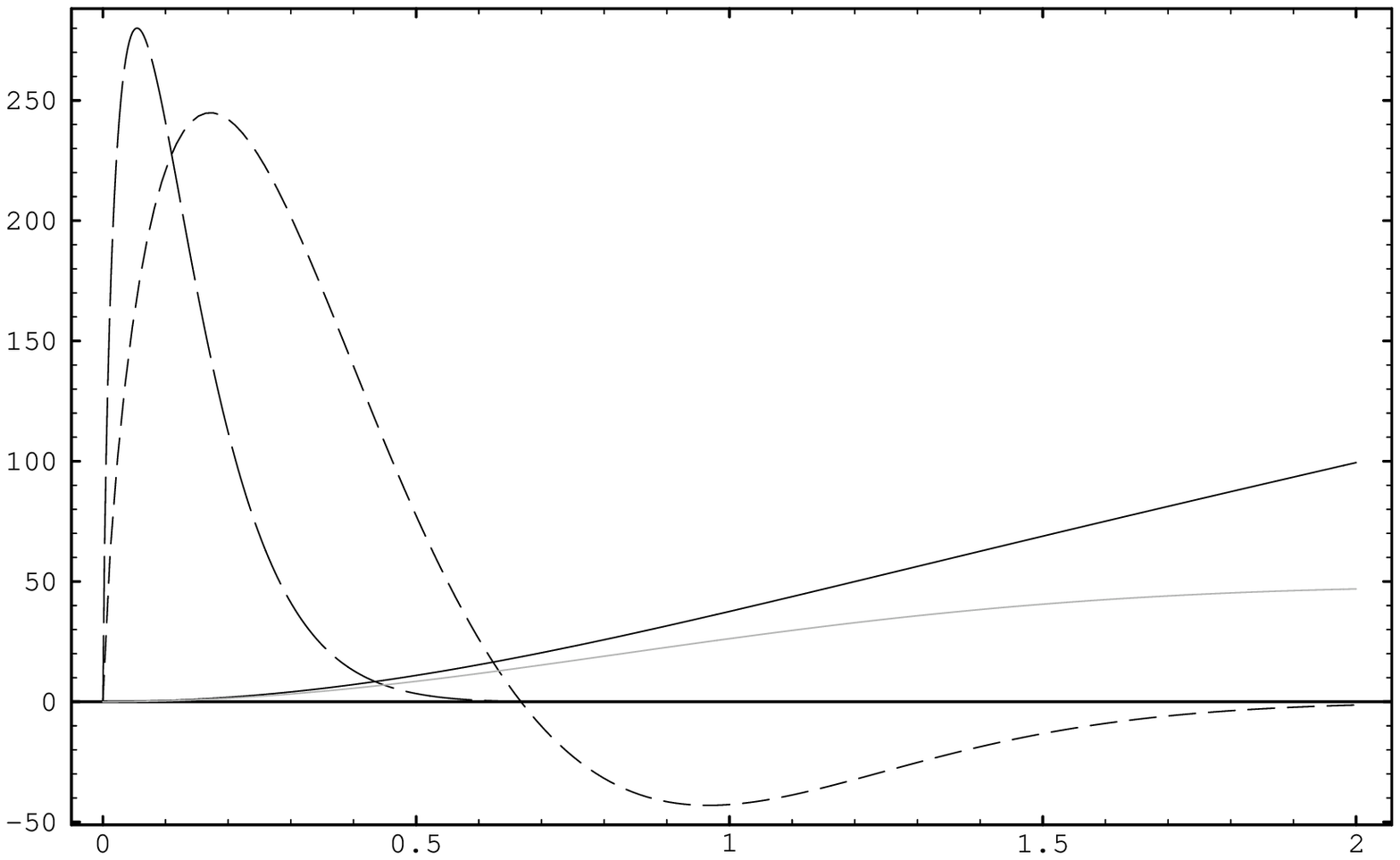}
\begin{picture}(0,0)
\put(8,5){\makebox(0,0){\LARGE $r[\!\!\fm]$}}
\put(-142,90){\makebox(145,10){
    \LARGE $r\ps_{V(\la)}^\dagger\ps_{\ga(Q^2,\la)}(r)[10^{-3}\fm^{-1}]$
    \hfill $J_p^{(0)}(1\!/\!2,r)[\rm mb]$}}
\put(-67,82){\makebox(0,0){\LARGE 2S-Longitudinal}}
\put(-95,60){\makebox(0,0){$Q^2\!=\!1\GeV^2$}}
\put(-112,84){\makebox(0,0){$Q^2\!=\!20\GeV^2$}}
\end{picture}
$$
\vspace*{-8mm}
$$
\figeps{140}{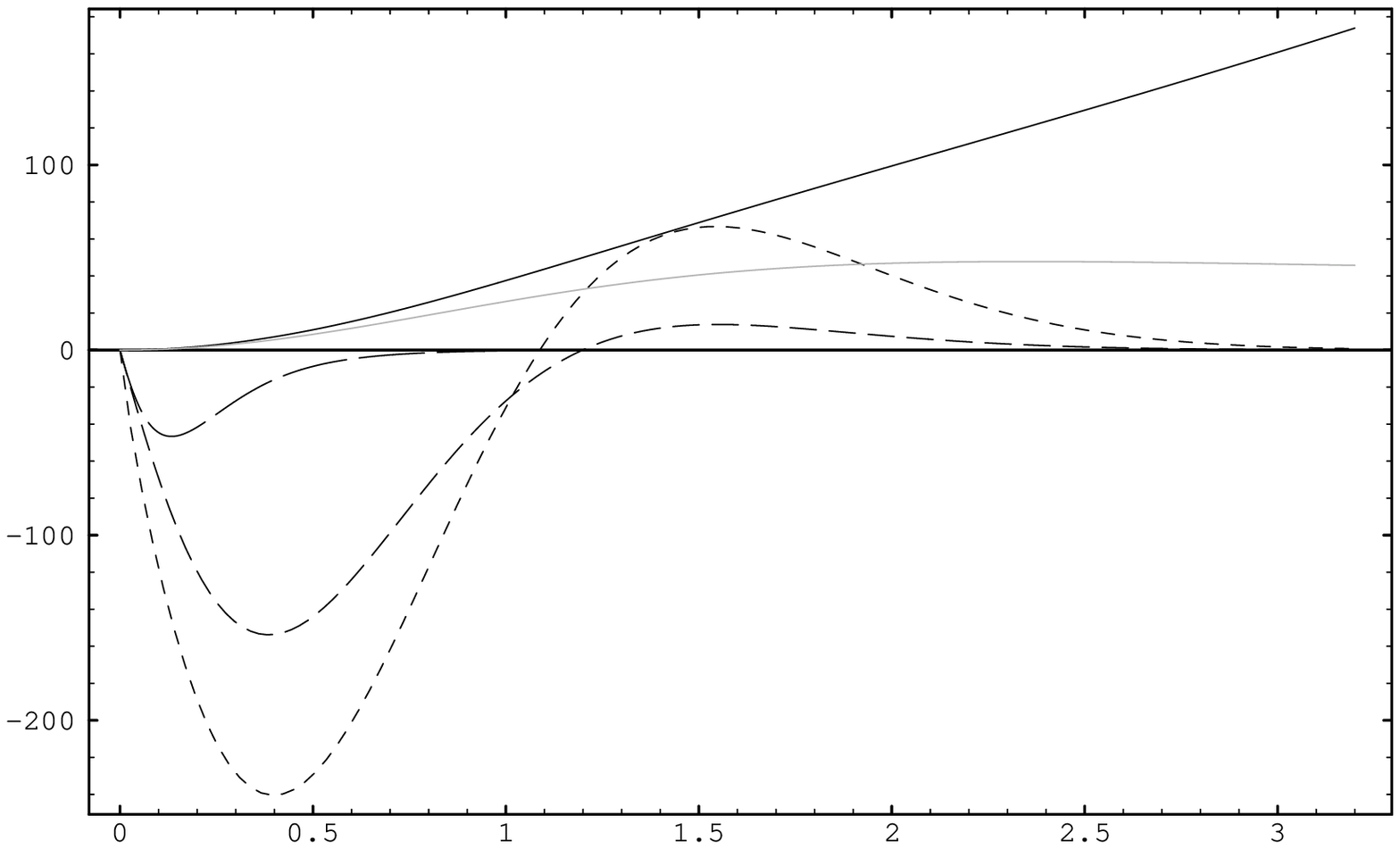}
\begin{picture}(0,0)
\put(8,5){\makebox(0,0){\LARGE $r[\!\!\fm]$}}
\put(-67,82){\makebox(0,0){\LARGE 2S-Transverse}}
\put(-101,12){\makebox(0,0){$Q^2\!=\!0$}}
\put(-111,38){\makebox(0,0){$Q^2\!=\!1\GeV^2$}}
\put(-105,47){\makebox(0,0){$Q^2\!=\!20\GeV^2$}}
\end{picture}
$$
\vspace*{-13mm}
\caption{ Dipole-proton total cross section $J_p^{(0)}$ and the effective
  overlap $r\ps_{V(\la)}^\dagger\ps_{\ga(Q^2,\la)}$ as function of the
  transverse dipole size $r$. The black lines are the function
  \mbox{$J_p^{(0)}(z\!=\!1\!/\!2, r)$} (Eq.~(\ref{sigma_dipole})), i.e. the
  total cross section of a dipole of fixed light-cone fraction $z\!=\!1\!/\!2$
  and transverse extension~$r$, averaged over all orientations, as a function
  of~$r$; the grey lines show the cross section of a completely abelian,
  non-confining theory.
  The leptoproduction amplitude is obtained by integration over the product
  of $J_p$ and the overlap function, which essentially (cf. Eq.~(\ref{overlap}))
  is the quantity shown for \mbox{$Q^2 \!=\! 0,\; 1$} and $20~\GeV^2$ as short,
  medium and long dashed curves, respectively. } \label{Fig:overlap2S}
\end{figure}
%

%
%
\begin{figure}
$$
\figeps{140}{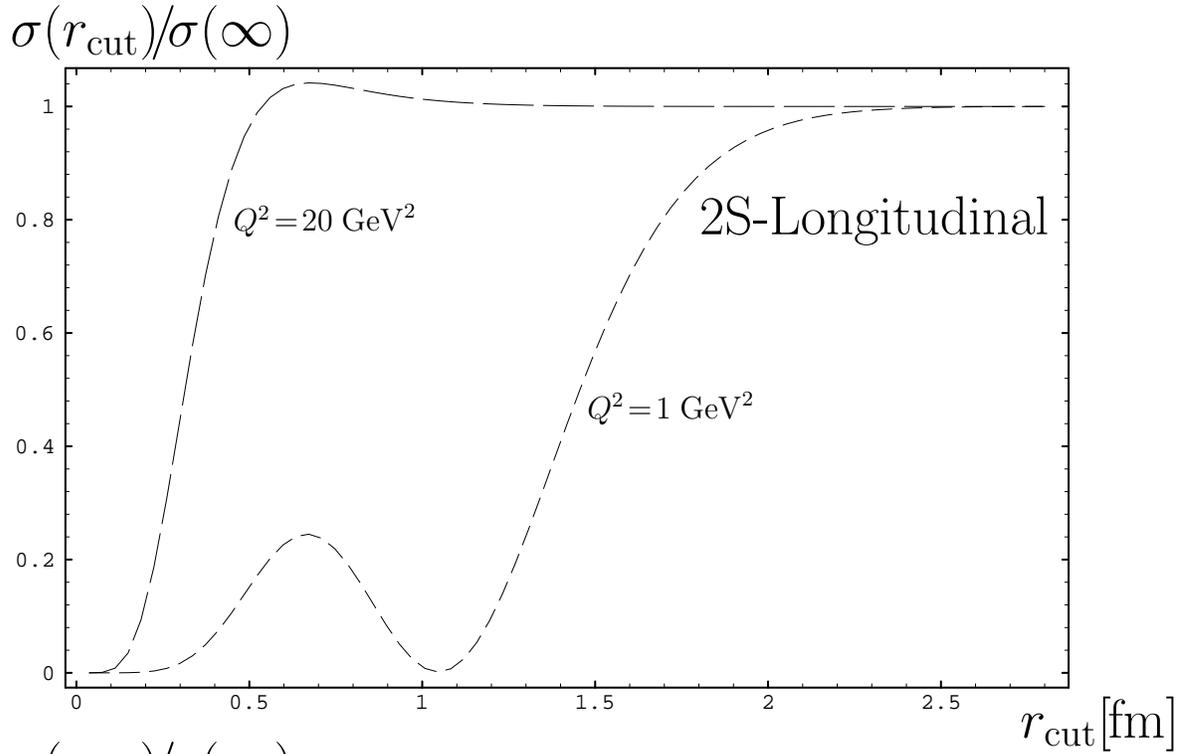}
\begin{picture}(0,0)
\put(4,2){\makebox(0,0){\LARGE $r_{\rm cut}[\!\!\fm]$}}
\put(-122,94){\makebox(0,0){\LARGE $\si(r_{\rm cut})\!/\!\si(\infty)$}}
\put(-26,69){\makebox(0,0){\LARGE 2S-Longitudinal}}
\put(-53,44){\makebox(0,0){$Q^2\!=\!1\GeV^2$}}
\put(-99,69){\makebox(0,0){$Q^2\!=\!20\GeV^2$}}
\end{picture}
$$
\vspace*{0mm}
$$
\figeps{140}{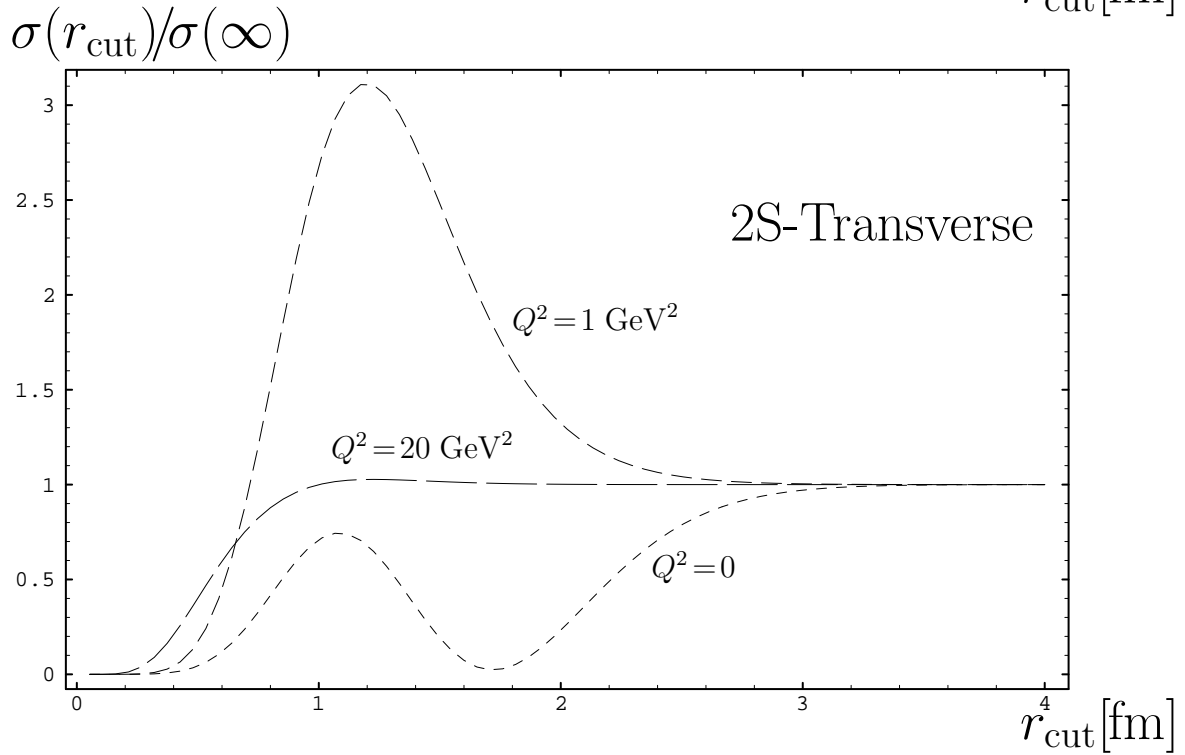}
\begin{picture}(0,0)
\put(4,2){\makebox(0,0){\LARGE $r_{\rm cut}[\!\!\fm]$}}
\put(-122,94){\makebox(0,0){\LARGE $\si(r_{\rm cut})\!/\!\si(\infty)$}}
\put(-25,69){\makebox(0,0){\LARGE 2S-Transverse}}
\put(-50,23){\makebox(0,0){$Q^2\!=\!0$}}
\put(-63,56){\makebox(0,0){$Q^2\!=\!1\GeV^2$}}
\put(-86,39){\makebox(0,0){$Q^2\!=\!20\GeV^2$}}
\end{picture}
$$
\vspace*{-7mm}
\caption{ Fraction of production cross sections due to dipole sizes smaller
  than $r_{\rm cut}$. The short, medium and long dashed curves refer to
  $Q^2\!=\!0,\; 1$ and $20\GeV^2$, respectively. Due to the node in the wave
  function of the $2S$-state (see Fig.~\ref{Fig:overlap2S}) the contribution
  of large dipole sizes is particularly important for small values of
  $Q^2$. } \label{Fig:Rcut2S}
\end{figure}

\subsection{Properties of the physical \mbox{\boldmath$\rho$}-,
  \mbox{\boldmath$\rho'$}- and \mbox{\boldmath$\rho''$}-states}

The mixing angle $\theta$ is determined by fitting the experimental
branching ratios 
\beqa \label{branchings}
X_1 &=& B_{e^+e^-} \, B_{\pi^+\pi^-} \;, \\
X_2 &=& B_{2\pi^+2\pi^-} / B_{\pi^+\pi^-} \;, \nn \\
X_3 &=& B_{\pi^+\pi^-} +  B_{2\pi^+2\pi^-} \nn
\end{eqnarray}
for the $\rh'$- and $\rh''$-resonances; it is calculated in App.~\ref{App:mixing}:
\beq
\theta=41.2^\circ \;.
\end{equation}
\begin{table}
\begin{center}
\begin{tabular}{|c||c|c|c|}
\hhline{|-||---|}
  & $\rho$ & $\rho'$ & $\rho''$ \\
\hhline{:=::===:}
$M_V[\!\!\GeV]$   & $0.7681\pm0.0013$  & $1.465\pm0.025$  & $1.700\pm0.020$ \\
$\Gamma_V^{tot}[\!\!\GeV]$
  & $0.1509\pm0.0030$ & $0.310\pm0.060$  & $0.235\pm0.050$ \\
$\Gamma_{V\rightarrow e^+e^-}[\!\!\GeV]$
  & $(6.77\pm0.32)\micro$
  & \mbox{\boldmath{$1.63\micro$}} & \mbox{\boldmath{$1.07\micro$}} \\
$f_V[\!\!\GeV]$   & 0.1526
  & \mbox{\boldmath{$-\,0.103$}}  & \mbox{\boldmath{$+\,0.0903$}} \\
$X_1$  & 4.48\mbox{$\times\!10^{-5}$}
  & 5.2\mbox{$\times\!10^{-7}$}  & 6\mbox{$\times\!10^{-7}$} \\
$X_2$  & 0  & 12.5  & 9.17 \\
$X_3$  & 1  & 0.8   & 0.8  \\
$B_{V\rightarrow\pi^+\pi^-}$    & 1  & 0.0593  & 0.0787 \\
$B_{V\rightarrow2\pi^+2\pi^-}$  & 0  & 0.741   & 0.721 \\ 
\hhline{|-||---|}
\end{tabular}
\end{center}
\caption[]{ Properties of the $\rho$-, $\rho'$- and $\rho''$-states.
  The couplings of $\rh'$ and $\rh''$ to the electromagnetic current (bold face)
  result from both the physical states taken as mixed states according to
  Eq.~(\ref{mixing}) and the state $|2S\rangle$ being normalized and orthogonal
  on $|1S\rangle$. For the masses, total and $\rh$-meson electromagnetic decay
  width see Ref.~ \cite{PDG}.
  The values $X_1$ and $X_2$ (see Eq.~(\ref{branchings})) for $\rh'$ and $\rh''$
  are taken from an analysis by Donnachie and Mirzaie~\cite{DM} of the
  $(\pi^+,\pi^-)$-mass spectra in photoproduction and
  \mbox{$e^+e^-$-annihilation}.
  Within the given accuracy we set $B_{\rh\rightarrow\pi^+\pi^-}\!=\!1$ and
  estimate the branching ratios of $\rh'$ and $\rh''$ in two or four charged pions
  to $80\%$.
  The last two lines summarize the branching ratios. } \label{Tabl:properties}
\end{table} 

In Table~\ref{Tabl:properties} we summarize the main properties of the three
physical states $\rh$, $\rh'$ and $\rh''$. There is considerable uncertainty
about the magnitude of $\Ga_{V\rightarrow e^+e^-}$ for the resonances $\rh'$ and
$\rho''$. The theoretical dilepton spectrum is in fair agreement with the data,
cf. Fig.~\ref{Fig:eebar2pis}.
In the $1.6\GeV$ region a destructive interference pattern shows up which fixes
the relative signs of the vector meson couplings $f_V$ as $(+,-,+)$. 
The signs of the decay constants of the $\rho'$ and $\rho''$ correspond to the
negative sign of the $2S$-wave function at the origin, cf. Eqs~(\ref{wfn2SL})
and (\ref{wfn2ST}), and the mixing angle $\theta\!=\!41.2^\circ$ in the first
quadrant.
In the same way as the dilepton width, $e^+e^-$-annihilation measures the short
range part of the wave functions; therefore the couplings $f_V$ essentially
determine both (see App.~\ref{App:llbar}).
At first sight the opposite pattern with a constructive interference in
photoproduction is puzzling, see Fig.~\ref{Fig:photo2pis}.

Because of the decay into $\pi a_1$ the $\rh'$- and $\rh''$-resonances are
considered as candidates for hybrid states in Ref.~\cite{Close}. On the other
hand the presence of an $\om \pi$-decay channel calls for a $2S$-component.
For the $\rh''$ the analysis of the decay channels does not demand a mixing,
but allows the presence of a hybrid component in the wave function.

\section{\hspace*{-.85em}. Diffractive cross sections} \label{Sect:diffraction}

The diffractive matrix elements for vector meson production are evaluated in the
specific model of the stochastic vacuum (MSV), see Refs~\cite{G1,G2,DFK}. 
One feature of this model is that the same mechanism which confines quarks also
induces a string-string interaction of colour singlet states which leads to a cross
section increasing with the $q\bar q$-dipole size roughly like $r^{1.5}$, when $r$
is in the interesting range of $1\,\fm \!<\! r \!<\! 2\, \fm$. 
In an alternative model of dipole-proton scattering, cf. Refs~\cite{KZ,NNN}, there
is a non-perturbative dipole-proton cross section which amounts to about half of
the total value for $0 \!\leq\! r \!\leq\! 2 \fm$.
The other half of the cross section comes from a perturbative two-gluon exchange
which saturates at $r\!=\!1 \fm$.
The difference between the two descriptions is most pronounced at distances
$1 \fm \!\leq\! r \!\leq\! 2 \fm$. 
Such large dipole sizes can only be tested with excited meson states like
the $\rho$-resonances. It is decisive to investigate the photo- and
leptoproduction of $\rh'$- and $\rh''$-mesons. These experiments hold the
key to find important long-range gluon fluctuations in diffraction which
are related to confinement in low-energy spectroscopy. 

The T-scattering amplitude is given by the integral over $z$ and $r$ of the wave
function overlap summed over quark helicities and multiplied with the
dipole-proton amplitude:
\beq \label{amplitude}
T^\la_V(s,t)= is \int \frac{dzd^2{\bf r}}{4\pi}\,
  \psi^\dagger_{V(\la)}\psi_{\ga(Q^2,\la)}(z,{\bf r})\, J_p(z,{\bf r},\Delta_T) \;,
\end{equation}
where the invariant momentum transfer squared $t\!=\!-\Delta_T^2$  (up to corrections
of the order $s^{-2}$, cf. Ref.~\cite{G1}) and the amplitude $J_p$ has the form:
\beq \label{jproton}
J_p(z,{\bf r},\Delta_T) = 2 \int_0^{\infty} bdb\, 2\pi J_0(\Delta_T b)
  \int \frac{dz_p d^2{\bf r}_p}{4\pi}\, |\psi_p(z_p,{\bf r}_p)|^2 
  J(b,z,{\bf r},z_p,{\bf r}_p) \;.
\end{equation}
The kernel $J(b,z,{\bf r},z_p,{\bf r}_p)$ is provided by the MSV and can be
understood as the interaction amplitude for the scattering of two colour dipoles,
where the second, with index "$p$", denotes a proton in the quark-diquark picture;
$b$ is the scattering impact parameter. This kernel as well as the profile function
\mbox{$J_p(z,{\bf r},\Delta_T)$} are the same as in previous work on moderate- and
high-$Q^2$ vector meson production. The Bessel function $J_0$ is obtained from the
angular integral in the Fourier transform. For a detailed discussion see
Ref.~\cite{G1}.

With Eq.~(\ref{amplitude}) as definition of the T-amplitude the differential cross
section with respect to $t$ writes
\beq
\frac{d\sigma^{\la}_V}{dt}(t) = \frac{1}{16\pi s^2}\, |T_V^{\la}(s,t)|^2 \;.
\end{equation}
Note, that the MSV evaluates $J_p$ in an eikonal approximation which causes the
T-amplitude to depend on $s$ only kinematically.
Integration over $t$ yield cross sections $\sigma^{\la}_V$ which are constant and
refer to a scattering energy $\sqrt{s}\!=\!20\GeV$ where the parameters of the model
are fixed (see discussion below).
For unpolarized photons the experimental data include transverse and longitudinal
contribution:
\beq
\si = \si^T + \ep\si^L \;,
\end{equation}
where the rate $\ep$ of longitudinally polarized photons depends on the lepton
scattering angle, the photon energy and virtuality and typically varies in the range
from $0.7$ to $1$, see Table~\ref{Tabl:cross_sections} and Ref.~\cite{G1}.

%
%
\begin{figure}
$$
\figeps{140}{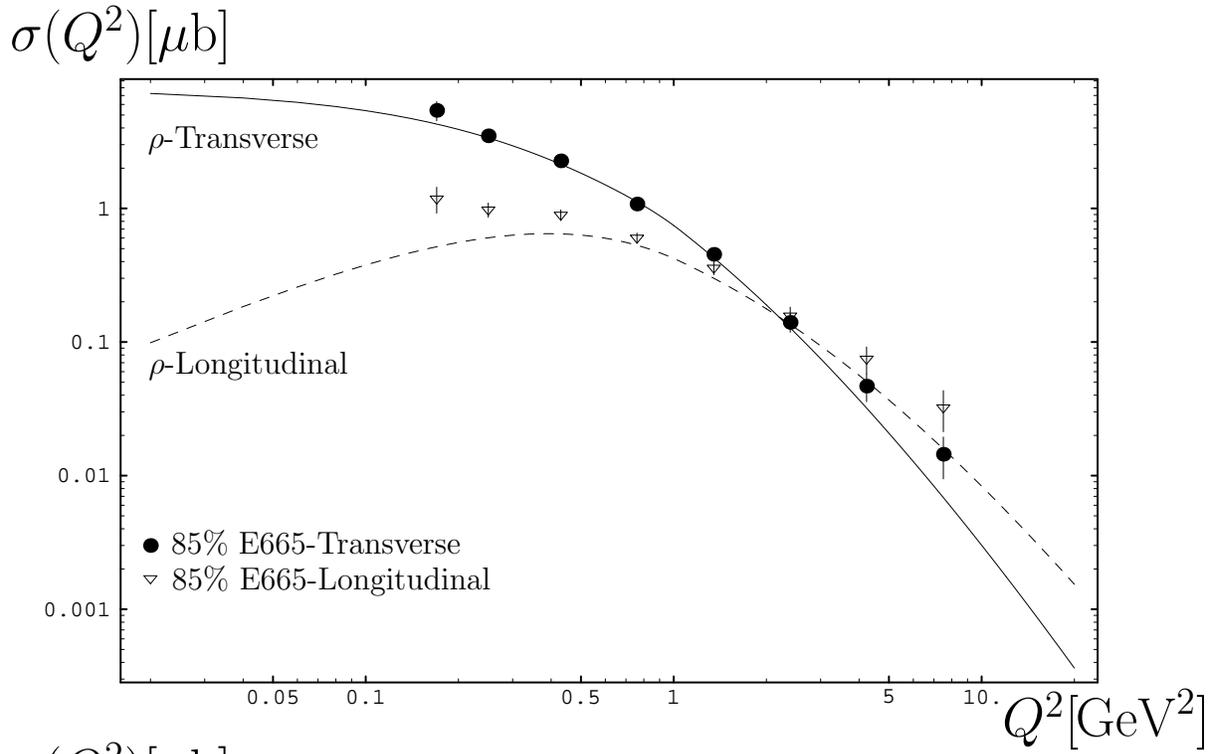}
\begin{picture}(0,0)
\put(1,3){\makebox(0,0){\LARGE $Q^2[\!\!\GeV^2]$}}
\put(-130,94){\makebox(0,0){\LARGE $\si(Q^2)[\rm \mu b]$}}
\put(-113,50){\makebox(0,0){$\rh$-Longitudinal}}
\put(-115,80){\makebox(0,0){$\rh$-Transverse}}
\put(-102,21.5){\makebox(0,0){85\% E665-Longitudinal}}
\put(-104,26.5){\makebox(0,0){85\% E665-Transverse}}
\end{picture}
$$
\vspace*{0mm}
$$
\figeps{140}{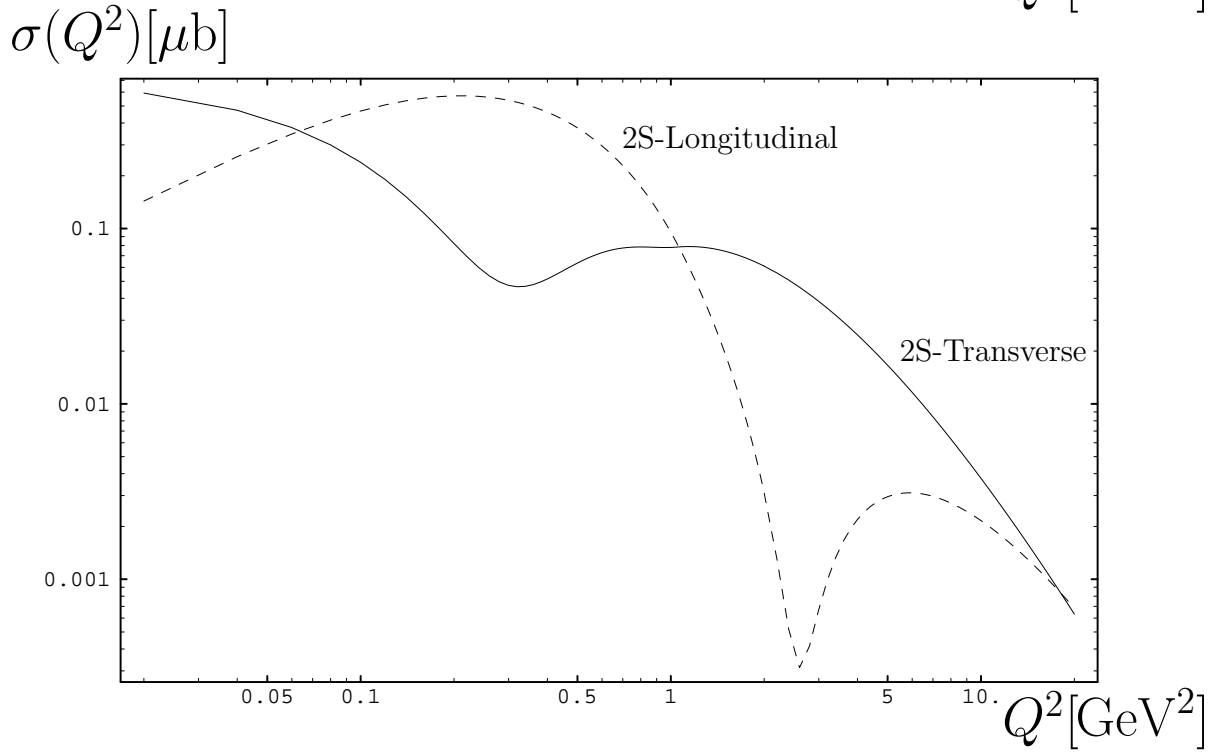}
\begin{picture}(0,0)
\put(1,3){\makebox(0,0){\LARGE $Q^2[\!\!\GeV^2]$}}
\put(-130,94){\makebox(0,0){\LARGE $\si(Q^2)[\rm \mu b]$}}
\put(-49,80){\makebox(0,0){2S-Longitudinal}}
\put(-14,52){\makebox(0,0){2S-Transverse}}
\end{picture}
$$
\caption[]{ Integrated elastic cross sections of the $\rho$-meson and the
  2S-state as a function of the photon virtuality $Q^2$. E665 \cite{E665_1}
  provides data for the $\rh$, cf. Table~\ref{Tabl:cross_sections}; we
  roughly estimate the pomeron contribution as 85\% of the measured cross
  section. } \label{Fig:sigma}
\end{figure}
%

%
%
\begin{sidewaystable}
\centering
\begin{minipage}{\linewidth}
\renewcommand{\thefootnote}{\thempfootnote}
\begin{tabular}{|f{4}||f{9}|c||f{8}|c||f{2}||f{8}|c|} \hline
\multicolumn{8}{|c|}{Integrated $\rh$-cross sections: theory vs experiment}
  \\ \hhline{:=:t:==:t:==:t:=:t:==:}
\multicolumn{1}{|c||}{$Q^2\;[\!\!\GeV^2]$}
  & \multicolumn{2}{c||}{$\si^T\;[\rm \mu b]$}
  & \multicolumn{2}{c||}{$\si^L\;[\rm \mu b]$} & \mbox{$\ep$}
  & \multicolumn{2}{c|}{$\si^T \!+\! \ep\si^L\;[\rm \mu b]$} \\
& \multicolumn{1}{c|}{th.\footnote{Pomeron contribution}} & \multicolumn{1}{c||}{exp.}  
  & \multicolumn{1}{c|}{ th.\footnotemark[\value{mpfootnote}] } & \multicolumn{1}{c||}{exp.}  
  & \multicolumn{1}{c||}{exp.} 
  & \multicolumn{1}{c|}{th.\footnotemark[\value{mpfootnote}] }
    & \multicolumn{1}{c|}{exp.} \\ \hhline{|-||--||--||-||--|}
0 & 7.86  & 9.4$\pm$1.1
              \footnote{For photon energies $20 \!<\! \nu \!<\! 70\GeV$, Ref.~\cite{Aston2}.}
  &   &   &   &   &   \\
0.17 & 4.28  & 6.37$\pm$0.89   & 0.517 & 1.39$\pm$0.26   & 0.76 & 4.67  & 7.42$\pm$0.91 \\
0.25 & 3.37  & 4.11$\pm$0.23   & 0.603 & 1.15$\pm$0.12   & 0.80 & 3.85  & 5.03$\pm$0.25 \\
0.43 & 2.14  & 2.67$\pm$0.13   & 0.645 & 1.051$\pm$0.081 & 0.81 & 2.66  & 3.52$\pm$0.15 \\
0.76 & 1.12  & 1.269$\pm$0.073 & 0.530 & 0.708$\pm$0.052 & 0.81 & 1.55  & 1.84$\pm$0.084 \\
1.35 & 0.426 & 0.533$\pm$0.045 & 0.300 & 0.422$\pm$0.040 & 0.81 & 0.669 & 0.875$\pm$0.055 \\
2.39 & 0.127 & 0.165$\pm$0.022 & 0.135 & 0.185$\pm$0.025 & 0.81 & 0.237 & 0.315$\pm$0.030 \\
2.5  & 115.\milli  & \non & 126.\milli & \non & 0.50 & 178.\milli & (170$\pm$31)\milli \\
3.5  & 51.6\milli  & \non & 71.7\milli & \non & 0.66 & 98.9\milli & (60$\pm$10)\milli \\
4.23 & 32.0\milli & (55$\pm$11)\milli & 50.7\milli & (88$\pm$17)\milli
  & 0.81 & 73.1\milli & (126$\pm$18)\milli \\
4.5  & 27.3\milli  & \non & 45.1\milli & \non & 0.66 & 57.1\milli & (65$\pm$11)\milli \\
5.5  & 16.1\milli  & \non & 30.4\milli & \non & 0.72 & 38.0\milli & (41$\pm$7)\milli \\
6.9  & 8.68\milli  & \non & 19.0\milli & \non & 0.76 & 23.1\milli & (23$\pm$3)\milli \\
7.51 & 6.85\milli & (17$\pm$5)\milli & 15.8\milli & (38$\pm$11)\milli
  & 0.81 & 19.7\milli & (47.8$\pm$10.2)\milli \\
8.8  & 4.36\milli  & \non & 11.1\milli & \non & 0.78 & 13.1\milli & (15$\pm$2)\milli \\
11.9 & 1.80\milli   & \non & 5.55\milli & \non & 0.82 & 6.35\milli & (5.8$\pm$0.9)\milli \\
16.9 & 0.617\milli & \non & 2.36\milli & \non & 0.81 & 2.53\milli & (2.6$\pm$0.7)\milli
  \\ \hhline{:=:b:==:b:==:b:=:b:==:}
\end{tabular}
\caption[]{ Theoretical cross sections for $\rh$-meson photo- and leptoproduction in
  comparison with data from NMC \cite{NMC} and E665 \cite{E665_1}, the latter with
  separate transverse and longitudinal polarizations. The experimental data contain a
  Regge contribution which at these energies can be estimated to about
  $15\%$. } \label{Tabl:cross_sections}
\end{minipage}
\end{sidewaystable}
\renewcommand{\thefootnote}{\arabic{footnote}}

In Fig.~\ref{Fig:overlap2S} we display the quantity
\beq \label{sigma_dipole}
J_p^{(0)}(z,r) :=
  \int_0^{2\pi} \frac{d\varph_{\bf r}}{2\pi}\,J_p(z,{\bf r},\Delta_T=0)
\end{equation} 
Due to the optical theorem it describes the total cross section of a dipole with
light-cone fraction $z$ and size $r$ (averaged over all its orientations) on a proton.
It depends only very slightly on $z$ and we display it for the central value
$z\!=\!1\!/\!2$. 
The grey lines show the contribution of a completely abelian model (which cannot yield
confinement), whereas the full lines represent the dipole-proton cross section as
evaluated in the MSV. The monotonous rise at large values of $r$ is a consequence of
a string-string interaction \cite{G1,DFK}. It depends crucially on the field strengths
correlators.
The input parameters have been fixed in order to obtain a consistent picture of the
slope of the $q\bar q$-confining potential, the numerical results for the correlators
from lattice simulations and proton-proton scattering at a scattering energy
$\sqrt{s}\!=\!20\GeV$, where hadron-hadron cross sections are approximately energy
independent.
All absolute cross sections calculated in the following refer thus to this energy.
For ratios of cross sections our results are also relevant at higher energies, since
change with $\sqrt{s}$ should affect numerator and denominator approximately in the same
way.

The second important input to the diffractive leptoproduction cross section are the
overlap matrix elements of the incoming photon with the outgoing vector meson. As has
already been pointed out in \cite{NNN} the node in the $2S$-state leads to a compensation
of contributions of large and small dipole sizes.
For our investigation this compensation is particularly interesting since it allows
very specific tests of the dipole cross section at large distances where the MSV makes
specific predictions.
In order to exhibit this effect we display in Fig.~\ref{Fig:overlap2S} the overlap
function which is for demonstration purpose integrated over $z$ and averaged over all
orientations
\beq \label{overlap}
r\ps_{V(\la)}^\dagger\ps_{\ga(Q^2,\la)}(r) :=
  \int \frac{dz}{4\pi}\; \int_0^{2\pi} \frac{d\varph_{\bf r}}{2\pi}\;
  |{\bf r}|\,\ps_{V(\la)}^\dagger\ps_{\ga(Q^2,\la)}(z,{\bf r}) \;,
\end{equation}
both for transverse and longitudinal photons and several values of $Q^2$. 

The T-amplitude $T^\la_V(s,t)$, cf. Eq.~\ref{amplitude}, can be estimated from
Fig.~\ref{Fig:overlap2S} by multiplying the dipole-proton cross section with the
overlap function and integrating over $r$.

The change of sign in the $2S$-wave function makes the T-amplitude very sensitive
to the behaviour of the dipole-proton cross section $J_p(z,r,\Delta_T)$ at larger
values of $r$.
Only its strong increase can overcome the negative contribution below the node and
lead to a positive sign of the imaginary part of $T^\la_V(s,t)$.
This will turn out to be crucial for the explanation of the different interference
patterns in photoproduction and $e^+e^-$-annihilation mentioned in Sect.~\ref{Sect:wfns}
and shown in Figs~\ref{Fig:eebar2pis} and \ref{Fig:photo2pis}.

The importance of the outer region, in particular for the $2S$-state, can also be seen
from Fig.~\ref{Fig:Rcut2S}:
There the cross section is calculated as function of an upper cut-off $r_{\rm cut}$
in the $r$-integration of Eq.~(\ref{amplitude}).
As can clearly be seen, in photoproduction the inner region of the overlap dominates
for $r \klgl 1.2\fm$, but compensation occurs at $r_{\rm cut} \!\cong\! 1.7\fm$
from the outer region, which contributes significantly to the $T$-amplitude up to
$r$-values of about $2.5\fm$.
By varying the photon-virtuality $Q^2$ one shifts the position of the node in the
overlap and thus the weight of the negative and positive contributions.
This is reflected in the strong \mbox{$Q^2$-dependence} of the
$\sigma(r_{\rm cut})$-curves in Fig.~\ref{Fig:Rcut2S} and the structured
\mbox{$Q^2$-dependence} of the transverse and longitudinal cross sections in
Fig.~\ref{Fig:sigma}. For the transverse cross section the outer positive region
dominates for $Q^2 \!\klgl\! 0.3\GeV^2$, where there is a dip in the $2S$-transverse
cross section; for the longitudinal one the dip is at $Q^2 \!\cong\! 2.5\GeV^2$.
For high $Q^2$-values only small dipole sizes $r$ contribute (early saturation in
Fig.~\ref{Fig:Rcut2S}) since the vector meson wave function suppresses the endpoint
values of the longitudinal momentum fraction $z$. 

In order to trace the experimental behaviour of the cross section as function of the
invariant mass M of the pions one needs the amplitudes of the $\rh'$- and
$\rh''$-resonances separately. 

Experimentally the $\pi^+\pi^-$- and $2\pi^+2\pi^-$-cross sections are measured.
Here it is essential to include the relevant branching ratios and the finite widths of
the resonances which in our case lead to a considerable reduction of the cross section
as compared to a zero-width approach. In detail we calculate for the final states
$f\!=\! \pi^+\pi^-$ and $2\pi^+2\pi^-$ the differential cross sections in respect to
their invariant mass M:
\beq \label{sigma_mass}
\frac{d\si^{f,\la}}{dM}
= \frac{2M}{16\pi s^2} \int dt \left| \sum_{V\!=\rh,\rh',\rh''}
  T^{\lambda}_V(s,t)\; \sqrt{\frac{M_V\Ga_V^{tot}}{\pi}}\;
  \frac{c_{V\!f}}{M^2 -M_V^2 + i M_V\Ga_V^{tot}}\;
  \sqrt{B_{V\rightarrow f}}\; \right|^2 \;.
\end{equation}
For the branching ratios $B_{V\rightarrow f}$ we refer to the discussion in
Table~\ref{Tabl:properties}; the $c_{V\!f}$ arise from proper normalization,
cf. Eqs~(\ref{width_norm}) and~(\ref{c_Vf}), and deviate from 1 on the
few-percent-level.

In the upper part of Fig.~\ref{Fig:sigma} we show the transverse and longitudinal
$\rh$-production cross section for values $0.02 \GeV^2 \!\le\! Q^2 \!\le\! 20 \GeV^2$
together with the data from NMC~\cite{NMC} and E665~\cite{E665_1}. 
The theoretical photoproduction cross section is \mbox{$\si_{\ga p} \!=\! 7.9 {\rm \mu b}$},
the experimental value \mbox{$9.4\pm1.1 {\rm \mu b}$}~\cite{Aston2}.
The experimental data contains also Reggeon, i.e. non-diffractive exchange which we
have not taken into account.
We may roughly
estimate the contribution at least for low $Q^2$ by compairing with the
Donnachie-Landshoff parametrization \cite{DoLa} of the total $\ga p$-cross section.
There at \mbox{$\sqrt{s} \!=\! 20\GeV$} the Reggeon contribution is 7 percent, hence we
estimate for the production cross section, i.e. the square of the amplitude, a
Reggeon contribution of 15 percent.
As constituent quark mass we use $m(Q^2)$ as it has been determined from the
vector current correlator and used in inclusive photoproduction, cf. Ref.~\cite{G2}.
At moderate $Q^2$ the E665 data are almost $20\!-\!30\%$ higher than our theoretical
calculations, but an extrapolation of the E665 data lies by about the same amount
above the NMC data, which we reproduce quite well. At $Q^2 \!>\! 1 \GeV^2$ the
theoretical cross sections are identical to the previously calculated
$\ga p \!\rightarrow\! \rh p$ cross sections, see Ref.~\cite{G1} and Eq.~(\ref{mQ2}).
Theory is confronted with the experimental data in more detail in
Table~\ref{Tabl:cross_sections}. 

The second part of Fig.~\ref{Fig:sigma} shows the integrated leptoproduction
cross section for the \mbox{$2S$-state}. The different node structure of the longitudinal
and transverse wave functions leads to slightly different behaviour. In the longitudinal
cross section there is a real zero at $Q^2\!\cong\!2.5\GeV^2$. In the transverse cross
section both helicity parts of the wave function, cf. Eq.~(\ref{wfn2ST}), contribute to
the overlap with the photon. The relativistic component with $L_z\!=\!1$ has its zero at
a different transverse separation than the nonrelavistic part with $L_z=0$ and aligned
quark spins.
Therefore the cross section has not a zero, but only a minimum at rather small
$Q^2 \!\cong\! 0.3\GeV^2$ and a plateau at \mbox{$Q^2 \!\cong\! 1 \GeV^2$}.
The magnitude of the cross section decrases in both cases since the photon wave function
shrinks in transverse extent at higher $Q^2$, and the inner negative parts of the excited
vector meson wave functions become dominant. Asymptotically the longitudinal cross section
dominates over the transverse by a power of $Q^2$; note, that for the $2S$-state we are not
in the asymptotic region even at $Q^2 \!=\! 20 \GeV^2$.

\begin{figure}
$$
\figeps{140}{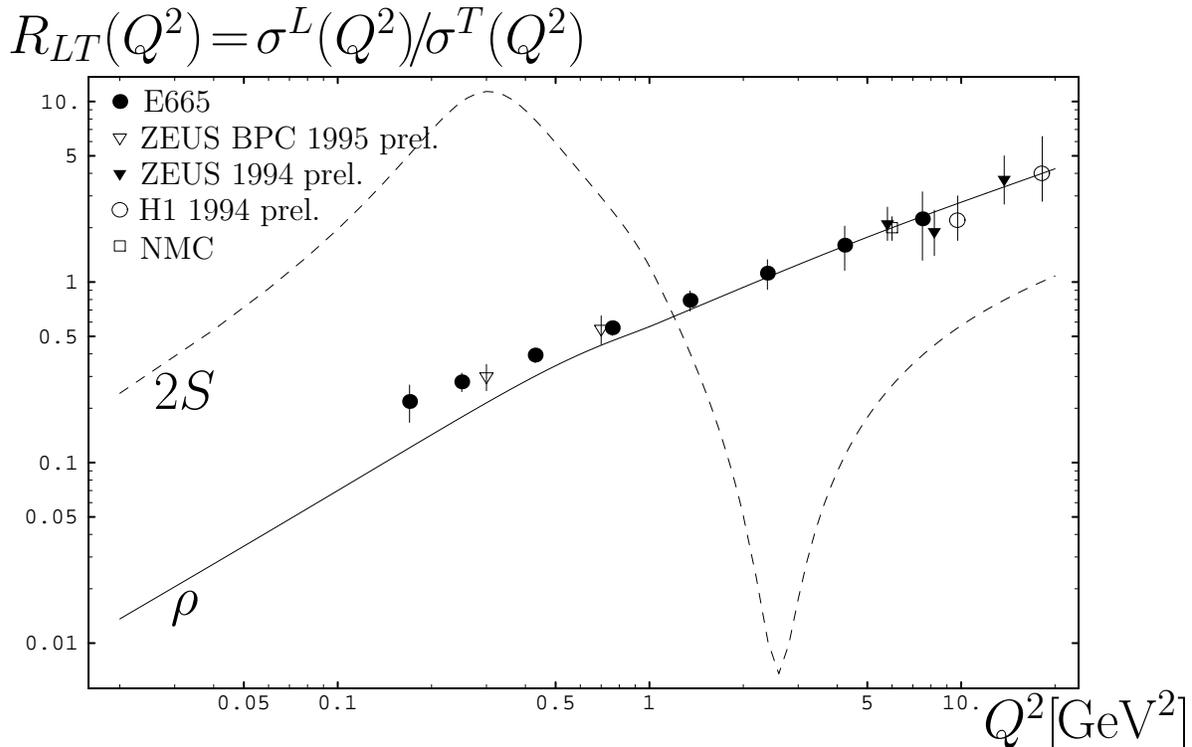}
\begin{picture}(0,0)
\put(1,3){\makebox(0,0){\LARGE $Q^2[\!\!\GeV^2]$}}
\put(-104,94){\makebox(0,0){\LARGE $R_{LT}(Q^2)\!=\!\si^L(Q^2)\!/\!\si^T(Q^2)$}}
\put(-119,18){\makebox(0,0){\LARGE $\rh$}}
\put(-119,47){\makebox(0,0){\LARGE $2S$}}
\put(-120,85.5){\makebox(0,0){E665}}
\put(-105,80.175){\makebox(0,0){ZEUS BPC 1995 prel.}}
\put(-110,75.5){\makebox(0,0){ZEUS 1994 prel.}}
\put(-113,70.625){\makebox(0,0){H1 1994 prel.}}
\put(-120,66){\makebox(0,0){NMC}}
\end{picture}
$$
\vspace*{-12mm} 
\caption{ Ratio of longitudinal over transverse integrated cross sections
  as function of $Q^2$ both for the $\rh$-meson (full) and the $2S$-state
  (dashed). There is only data for $\rh$-production. } \label{Fig:R_LT}
\end{figure}

In Fig.~\ref{Fig:R_LT} we show the ratio
\beq \label{R_LT}
R_{LT} = \sigma_L/\sigma_T
\end{equation}
of longitudinal to transverse cross sections for the $\rh$- and $2S$-states including all
$Q^2$ virtualties up to $20\GeV^2$.
For the $\rh$-meson the rapid rise of $R_{LT}$ is confirmed quite well.
We remark that an analysis of colour transparency in nuclei \cite{Adams} should include
a rapidly increasing ratio of longitudinal to transverse cross sections.

%
%

\newpage
%
%
\begin{sidewaysfigure}
\vspace*{4mm}
\setlength{\unitlength}{1mm}
\begin{picture}(235,145.5)
\put(-2,72.75){ \makebox(117.4,72.62){\figeps{117}{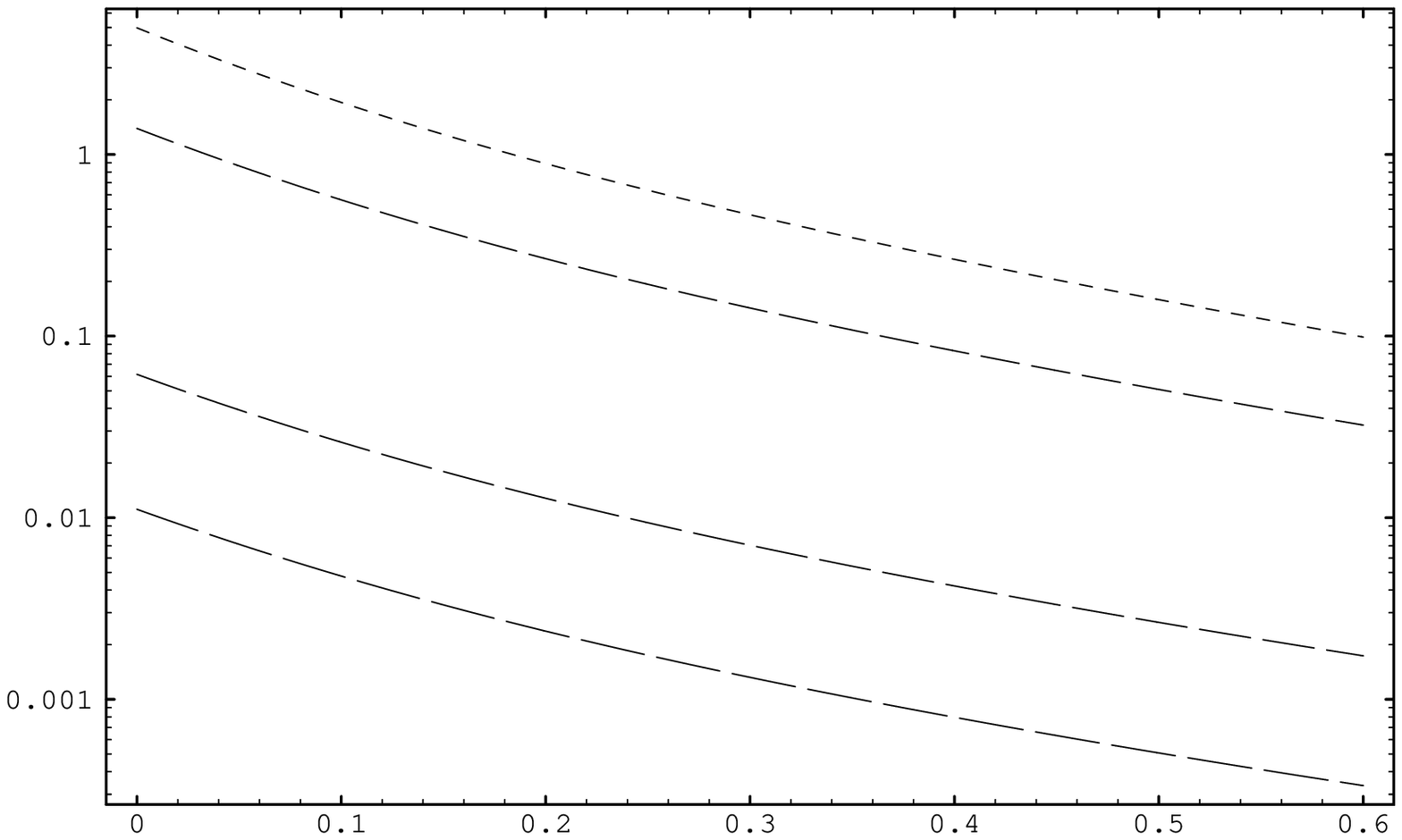}} }
  \put(28,149.75){\makebox(0,0){
      \LARGE $d\si\!/\!dt[\mu b \!\times\!\!\!\GeV^{-2}]$}}
  \put(95,140.75){\makebox(0,0){\LARGE $\rh$-Longitudinal}}
  \put(45,139.75){\makebox(0,0){$Q^2\!=\!1\!/\!4\GeV^2$}}
  \put(23,124.75){\makebox(0,0){$Q^2\!=\!2\GeV^2$}}
  \put(45,110.75){\makebox(0,0){$Q^2\!=\!10\GeV^2$}}
  \put(24,92.25){\makebox(0,0){$Q^2\!=\!20\GeV^2$}}
%
\put(117.5,72.75){ \makebox(117.4,72.62){\figeps{117}{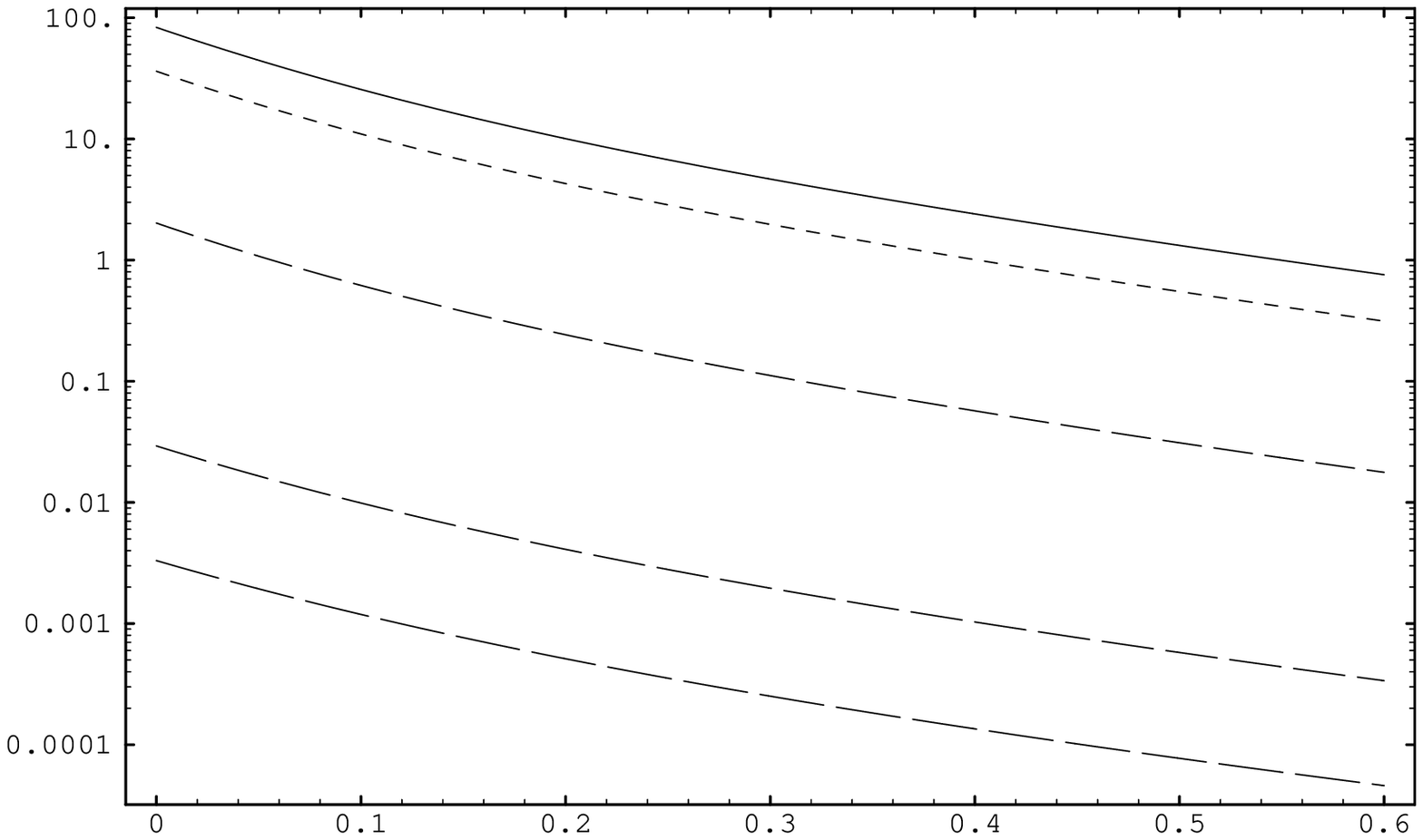}} }
  \put(216.5,139.75){\makebox(0,0){\LARGE $\rh$-Transverse}}
  \put(171,136.75){\makebox(0,0){$Q^2\!=\!0$}}
  \put(177.5,122.75){\makebox(0,0){$Q^2\!=\!1\!/\!4\GeV^2$}}
  \put(142.5,117.75){\makebox(0,0){$Q^2\!=\!2\GeV^2$}}
  \put(177.5,102.75){\makebox(0,0){$Q^2\!=\!10\GeV^2$}}
  \put(143.5,89.75){\makebox(0,0){$Q^2\!=\!20\GeV^2$}}
%
\put(-4.65,0){ \makebox(117.4,72.62){\figeps{122.3}{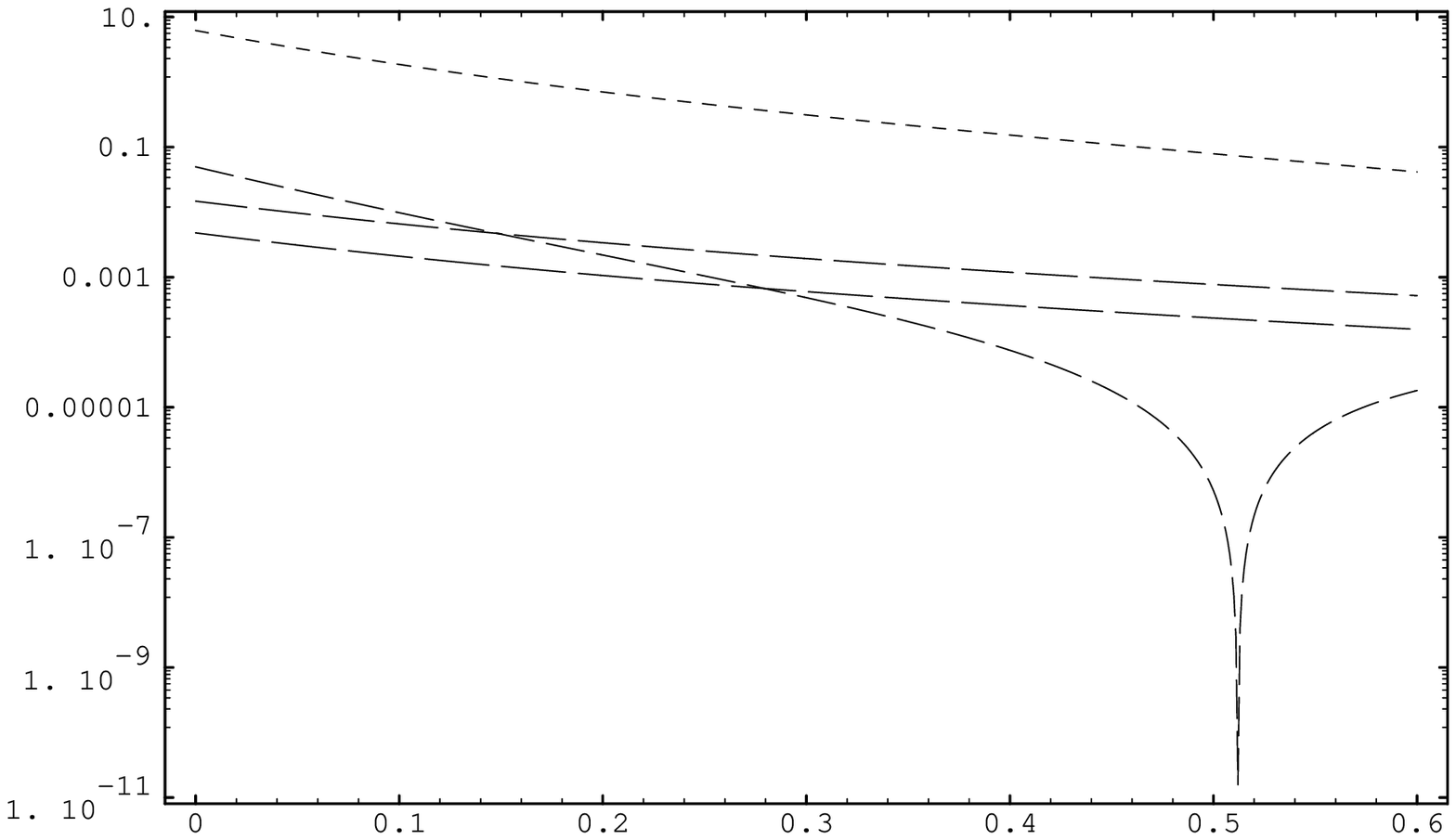}} }
  \put(93,67){\makebox(0,0){\LARGE 2S-Longitudinal}}
  \put(57,68){\makebox(0,0){$Q^2\!=\!1\!/\!4\GeV^2$}}
  \put(28,61.5){\makebox(0,0){$Q^2\!=\!2\GeV^2$}}
  \put(57,55.5){\makebox(0,0){$Q^2\!=\!10\GeV^2$}}
  \put(29,48){\makebox(0,0){$Q^2\!=\!20\GeV^2$}}
%
\put(116.8,0){ \makebox(117.4,72.62){\figeps{118.5}{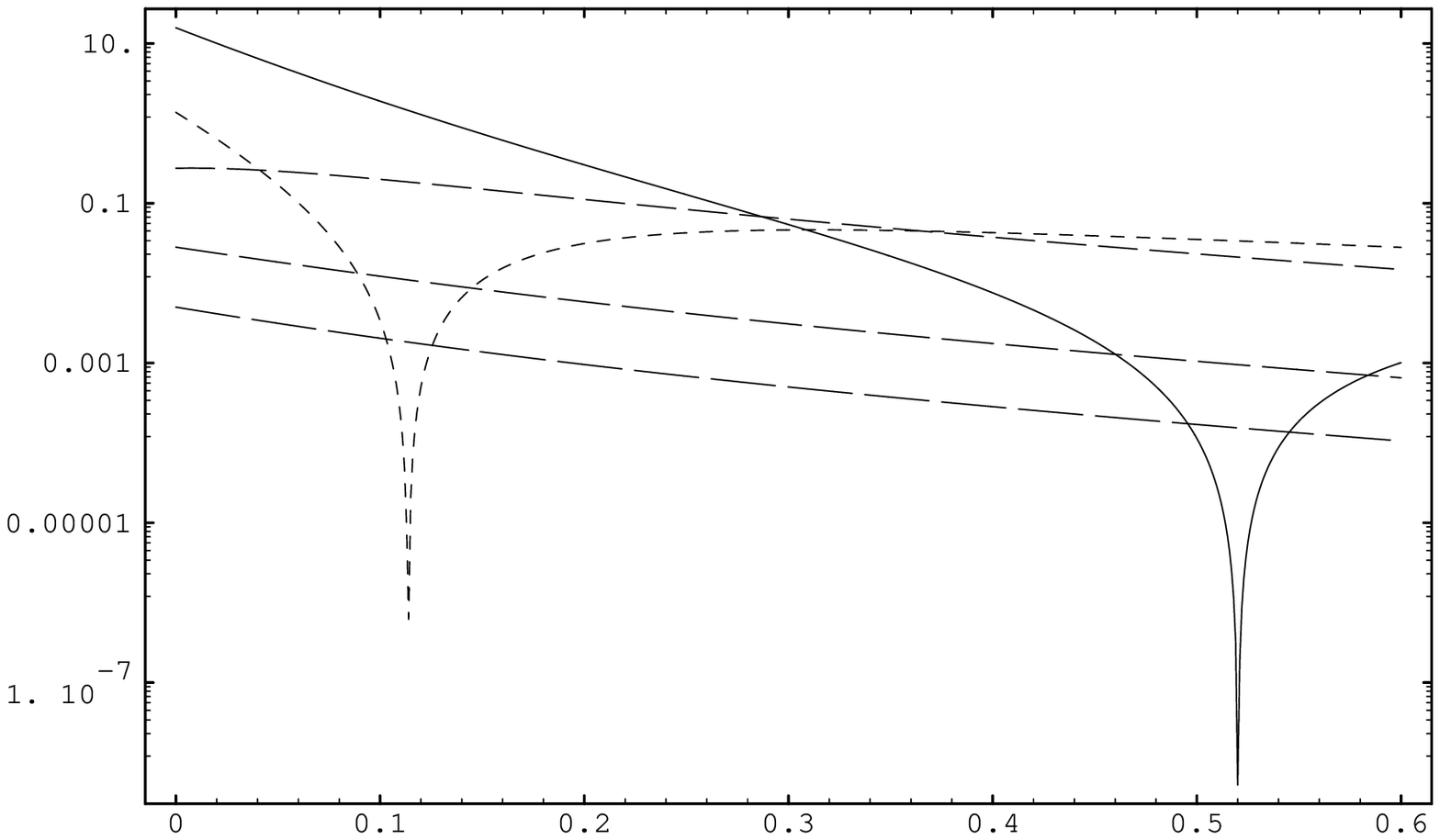}} }
  \put(223.5,-0.5){\makebox(0,0){\LARGE $-t[\!\!\GeV^2]$}}
  \put(214.5,68){\makebox(0,0){\LARGE 2S-Transverse}}
  \put(212.5,11){\makebox(0,0){$Q^2\!=\!0$}}
  \put(144.5,17){\makebox(0,0){$Q^2\!=\!1\!/\!4\GeV^2$}}
  \put(222.5,48){\makebox(0,0){$Q^2\!=\!2\GeV^2$}}
  \put(177.5,50){\makebox(0,0){$Q^2\!=\!10\GeV^2$}}
  \put(177.5,37.5){\makebox(0,0){$Q^2\!=\!20\GeV^2$}}
%
\end{picture}
\vspace*{-7mm}
\caption[]{ Differential cross section as a function of $-t$ for the
  $\rh$-meson and the $2S$-state (upper and lower plots) for both
  longitudinal and transverse polarization (left and right).
  The curves with increasing dash sizes refer to 
  \mbox{$Q^2\!=\!0,\; 1\!/\!4,\; 2,\; 10,\; 20\GeV$}.
  For the \mbox{$2S$-state} the node in the wave function has a strong
  influence on the $t$-dependence. } \label{Fig:dsdt}
\end{sidewaysfigure}

In Fig.~\ref{Fig:dsdt} the respective differential cross sections are shown.
For the $1S$-meson production the longitudinal and transverse differential
cross sections follow roughly exponential behaviour with a slight upward
curvature at larger $-t$ values.
In comparison the $2S$-state produces sharp dips in the differential cross
sections which occur at the same $Q^2$ where the integrated cross sections
have minima. The occurence of the dips is a consequence of the node in the
wave function, the exact location of these minima is highly parametrization
dependent. At these $Q^2$-values where the minima occur the cross section is
much faster falling off than in general. For an experiment where the
superposition of longitudinal and transverse cross sections will be measured,
these sharper fall-offs may be a good signal for interesting physics.

%
%
\begin{figure}
$$
\figps{140}{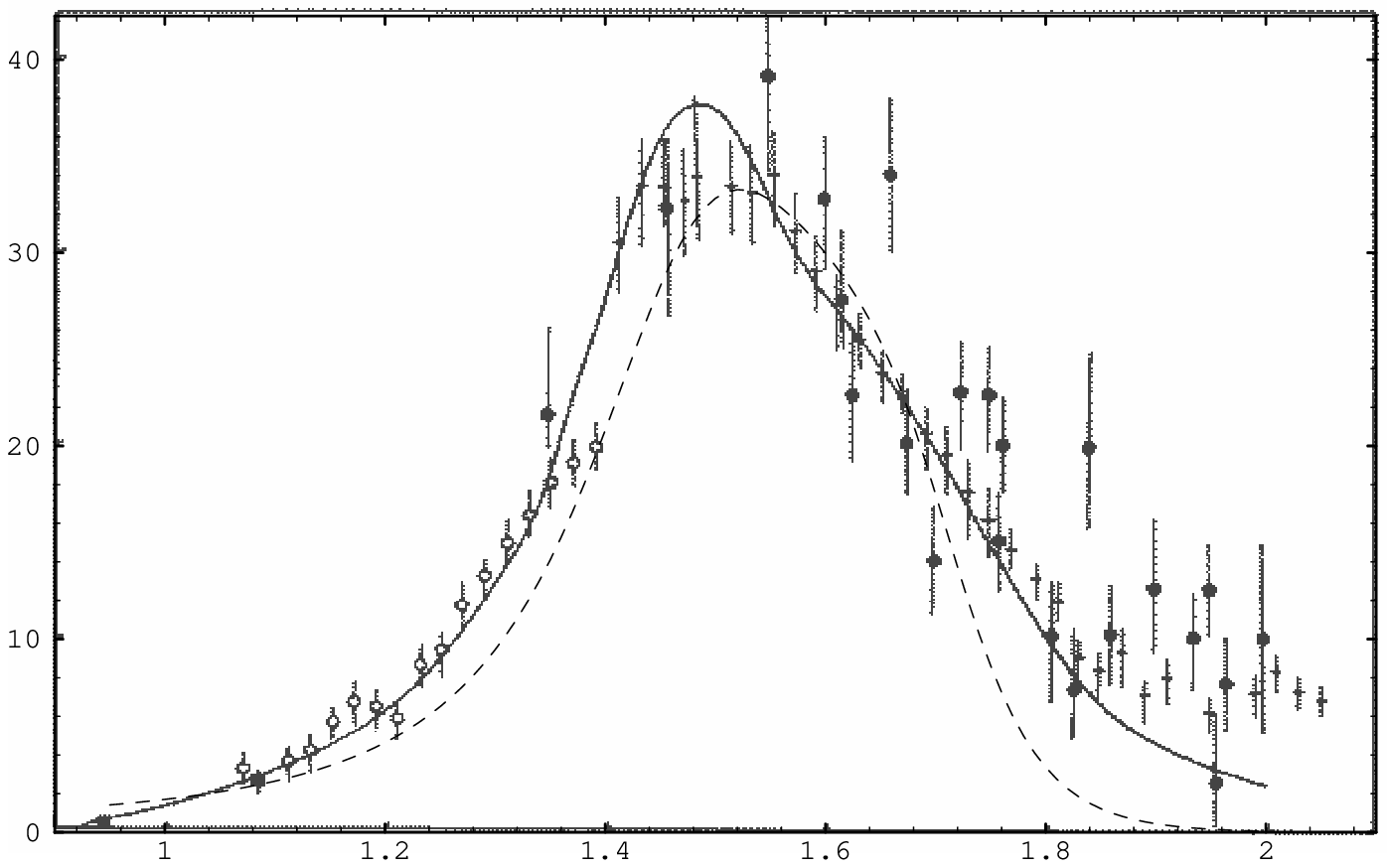}
\begin{picture}(0,0)
\put(3,3){\makebox(0,0){\LARGE $M[\!\!\GeV]$}}
\put(-140,91){\makebox(140,0){\LARGE $\si(M)[nb]$
    \hfill $e^+e^-\rightarrow2\pi^+2\pi^-$}}
\end{picture}
$$
\vspace*{-11mm}
\caption[]{ Mass spectrum of $e^+e^-$-annihilation into $2\pi^+2\pi^-$.
  The full line is a fit by Donnachie and Mirzaie~\cite{DM}. The dashed
  line is the result of the parametrization used in this paper
  (see~Table~\ref{Tabl:properties} and App.~\ref{App:llbar}). } \label{Fig:eebar4pis}
\end{figure}
%

%
%
\begin{figure}
$$
\figeps{140}{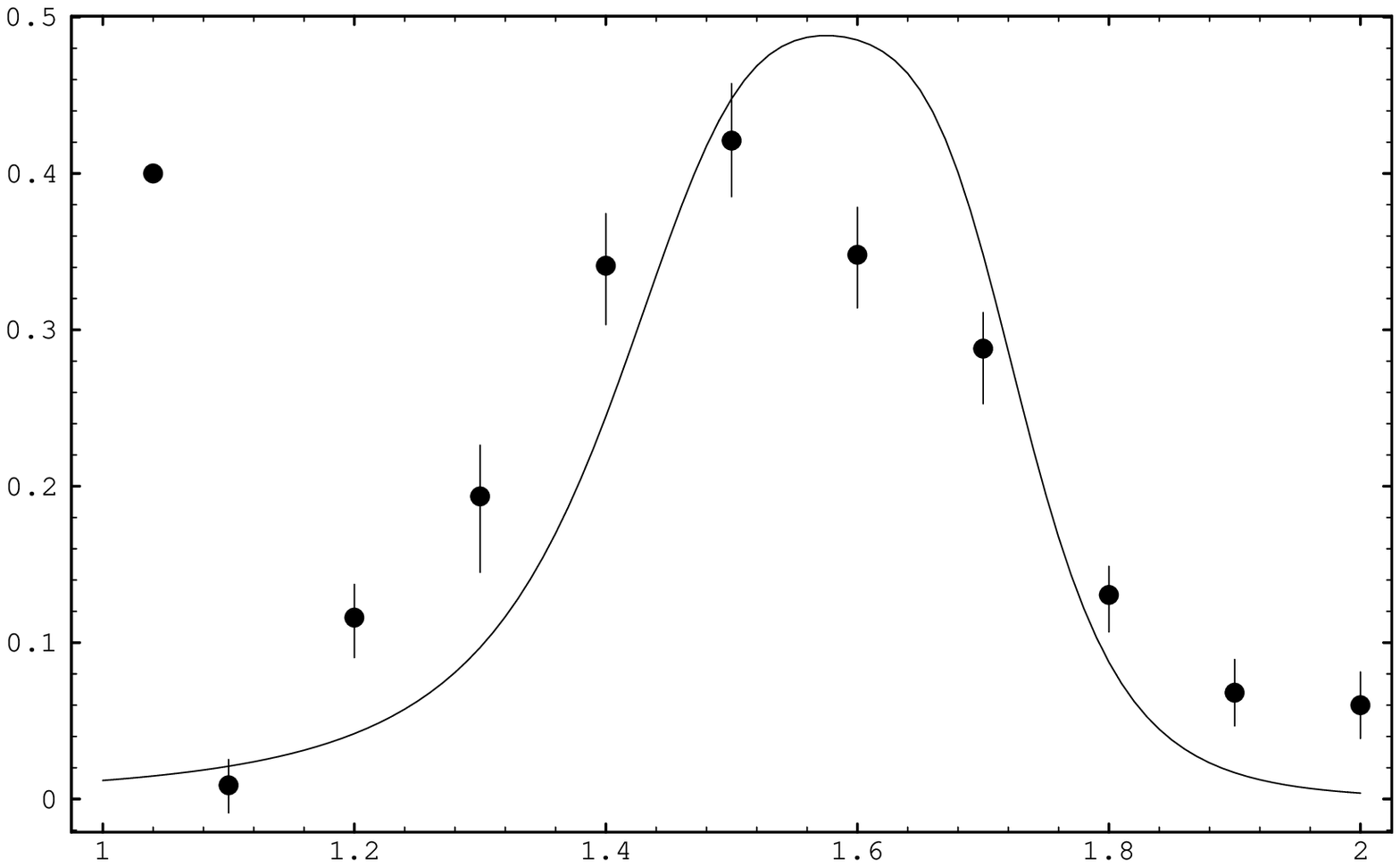}
\begin{picture}(0,0)
\put(3,-0.5){\makebox(0,0){\LARGE $M[\!\!\GeV]$}}
\put(-140,94.5){\makebox(145,0){\LARGE $d\si\!/\!dM(M)[\mu b\!/\!\!\GeV]$
    \hfill $\ga p\rightarrow2\pi^+2\pi^-p$ }}
\put(-118.5,73.3){\makebox(0,0){50\%}}
\put(-108.5,68.3){\makebox(0,0){Aston et al. \cite{Aston2}}}
\end{picture}
$$
\vspace*{-8mm}
\caption[]{ Our result for the mass spectrum of photoproduced
  $2\pi^+2\pi^-$. The data are from Ref.~\cite{Aston2}, scaled with a
  factor $0.5$\;. } \label{Fig:photo4pis}
\end{figure}

We revisit now the $\pi^+\pi^-$- and $2\pi^+2\pi^-$-production experiments
across the $1\!-\!2\GeV$ mass region. The most interesting result of the
experiments is the different interference pattern in $e^+e^-$-annihilation
compared to photoproduction. In our convention the $2S$-wave function is
negative relative to the ground state wave function at the origin so that the
interference pattern determines the mixing angle to be in the first quadrant
which gives opposite signs for the $\rh'$- and $\rh''$-annihilation amplitudes,
i.e. the $\rh$-, $\rh'$- and $\rh''$-annihilation amplitudes have the relative
signs $(+,-,+)$.
In Fig.~\ref{Fig:eebar2pis} the dashed curve shows the theoretical
$\pi^+\pi^-$-mass distribution in $e^+e^-$-annihilation according to the
parametrization for $\rh$, $\rh'$ and $\rh''$ used in this paper (see
Table~\ref{Tabl:properties} and App.~\ref{App:llbar}). The data are from Orsay
and Novosibirsk, Refs~\cite{Ba,Bi}; its main feature is the destructive
interference slightly above $1.5\GeV$ which is correctly reproduced with the
mixing angle $\th \!=\! 41.2^\circ$. Our parametrization also gives a sizeable
$2\pi^+2\pi^-$-cross section of $40$ nb in this range, see
Fig.~\ref{Fig:eebar4pis}.

For photoproduction, however, cf. Fig.~\ref{Fig:photo2pis}, experimental cross
section from SLAC \cite{Aston1} show a small enhancement near the same energy.
More recent data \cite{E687} even point to the possibility of $3S$-production.
The different interference of the photoproduction amplitudes $T_V$, obeying the
sign pattern $(+,+,-)$, comes from the dipole character of the cross section
which favours the large-$r$ part of the vector meson wave function more than the
short range part which is important for the $e^+e^-$-coupling.
The theoretical $2\pi^+2\pi^-$-photoproduction cross section is shown in
Fig.~\ref{Fig:photo4pis}. It is experimentally very demanding to subtract the
background in order to identify the resonating contribution.

\begin{figure}
$$
\figeps{140}{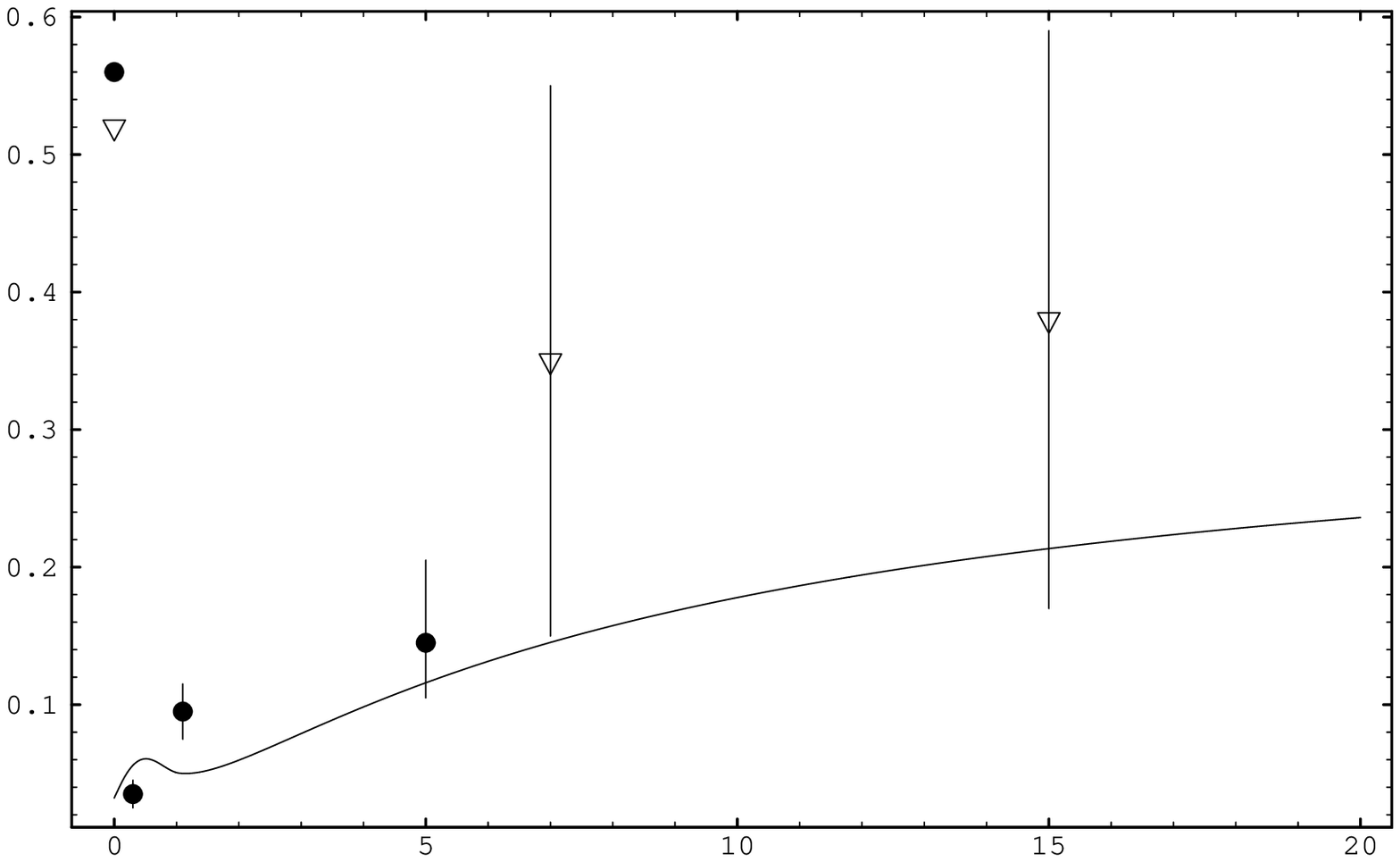}
\begin{picture}(0,0)
\put(1,0){\makebox(0,0){\LARGE $Q^2[\!\!\GeV^2]$}}
\put(-128,94){\makebox(0,0){\LARGE $R_\pi(Q^2)$}}
\put(-121,83){\makebox(0,0){E665}}
\put(-118,77){\makebox(0,0){H1 prel.}}
\end{picture}
$$
\vspace*{-9mm}
\caption[]{ Ratio $R_\pi$ of
  $2\pi^+2\pi^-$-production via $\rh'$ and $\rh''$
  over $\pi^+\pi^-$-production via $\rh$
  as function of $Q^2$, cf. Eq.~(\ref{R_pis}). Our curve is for
  to $\ep\!=\!1$. Experimental points are taken from
  Refs~\cite{E665_2} and~\cite{H1}. } \label{Fig:R_pis}
\end{figure}

Therefore it might be more realistic to look for the strong variation in the
observables with $Q^2$ than for the absolute values.
It should be noted that the strong variations of the cross sections with $Q^2$ in
Figs~\ref{Fig:sigma}, ~\ref{Fig:R_LT} and \ref{Fig:R_pis} are a clear prediction of
our model, the exact positions of the dips, however, depend crucially on the exact
position of the node in the $2S$-state.

%
%
%

Experimentally also accessible is the ratio
of $2\pi^+2\pi^-$-production via $\rh'$ and $\rh''$
over $\pi^+\pi^-$-production via $\rh$:
\beq \label{R_pis}
R_\pi = \frac{ \left. \si_{(\rh'\!\rh'')}^{f,T}
                + \ep \si_{(\rh'\!\rh'')}^{f,L} \right|_{f=2\pi^+2\pi^-}
            }{ \left. \si_{(\rh)}^{f',T}
                + \ep \si_{(\rh)}^{f',L} \right|_{f'=\pi^+\pi^-} } \;,
\end{equation}
where
\beqa
\si_{(\rh)}^{f,\la} &=& \int_{s_f}^\infty dM^2\, 
  \frac{1}{16\pi s^2} \int dt \left|\;
  T^{\lambda}_\rh(s,t)\; \sqrt{\frac{M_\rh\Ga_\rh^{tot}}{\pi}}\;
  \frac{c_{\rh f}}{M^2 -M_\rh^2 + i M_\rh\Ga_\rh^{tot}}\;
  \sqrt{B_{\rh\rightarrow f}}\; \right|^2 \;, \\
\si_{(\rh'\!\rh'')}^{f,\la} &=& \int_{s_f}^\infty dM^2\, 
  \frac{1}{16\pi s^2} \int dt \left| \sum_{V\!=\rh',\rh''}
  T^{\lambda}_V(s,t)\; \sqrt{\frac{M_V\Ga_V^{tot}}{\pi}}\;
  \frac{c_{V\!f}}{M^2 -M_V^2 + i M_V\Ga_V^{tot}}\;
  \sqrt{B_{V\rightarrow f}}\; \right|^2 \;. \nn
\end{eqnarray}
The interference between the $\rh'$ and $\rh''$ reduces the result compared
to the zero-width approximation to $47\!\pm\!18\,\%$, where the uncertainty
is due to the uncertainty in the widths of the resonances
(see Table~\ref{Tabl:properties}).

In Fig.~\ref{Fig:R_pis} we show $R_\pi$ for $\ep\!=\!1$:
After a small fluctuation near $Q^2 \!=\! 1\GeV^2$ it increases continuously.
The structure at small $Q^2$ comes from the conspiracy of the longitudinal and
transverse parts in the ratio; due to our lack of knowledge of the exact position
of the nodes there may be only a continuous rise in $R_\pi$ instead of a
structured fluctuation.
Especially the transverse cross section is very sensitive to the shape of the
$2S$-state, since the large transverse photon extends well across the node.
The longitudinal photon even at small $Q^2$ mainly tests the inner negative part
of the state.
At high $Q^2$ our calculation of $R_\pi$ should truely reflect physics at work
and indeed is in qualitative agreement with preliminary data from H1 \cite{H1}
which still have large error bars.

\section{\hspace*{-.85em}. Discussion and summary} \label{Sect:discussion}

In conclusion we have presented a realistic calculation for photoproduction which is
based on a description of the $\rh'$- and $\rh''$-mesons as mixed quark-antiquark
$2S$-states with some inert residual component.
The decay characteristics of the $\rh'$ point towards a sizeable hybrid admixture
which may exist also for the $\rh''$.
With our ansatz the different interference patterns in $e^+e^-$-annihilation and
photoproduction of two charged pions induce a mixing angle which implies that the
$\rh'$ and $\rh''$ are about one half a quark-antiquark $2S$-state and one half
hybrid or $2D$-excitation.
In this paper we give further evidence for the validity of the picture of diffraction
as scattering of colour neutral states due to long-range gluon fluctuations.
The large vector meson excited states test favourably our picture of a dipole-proton
cross section increasing with the quark-antiquark transverse distance $r$ due to
string-string interactions which emerge from the model of the stochastic vacuum as a
typical consequence of non-perturbative QCD.
Especially transverse photoproduction has a matrix element where the elementary
dipole-proton cross section is sampled between $1$ and $2\fm$.
It is the dipole-proton cross section in that range which explains the markedly
different interference patterns for $e^+e^-$-annihilation and photoproduction.
If it turns out that with increasing energy the excitation of the residual hybrid
state becomes more important, we would see some indication for a perturbative gluonic
component in the photon wave function which has matrix elements with the intrinsic glue
in the hybrid. This would at the same time open up a window to the world of nonexotic
hybrids and give us more insight into the importance of perturbative physics in
diffraction.
We have calculated all diffractive cross sections at $\sqrt{s} \!=\! 20\GeV$, but argued that
the calculated ratios as $\si_L/\si_T$ are also valid at higher energies. This is seen
in the good agreement with HERA-data~\cite{H1}.
A big challenge remains to combine this picture of long range string-string
interactions due to the stochastic vacuum with short range perturbative gluon
fluctuations in order to understand the energy dependence of diffraction.

\renewcommand{\thesubsection}{}
\subsection{ Acknowlegments } 

\noindent
We thank Sandi Donnachie for many illuminating discussions during his stay
in Heidelberg and E.L.~Gubankova for participation in the early stage of the
work.

%
%
%
\appendix
\renewcommand{\thesection}{\Alph{section}}
\renewcommand{\theequation}{\thesection \arabic{equation}}
\renewcommand{\thefootnote}{\Alph{section}\hspace*{-1pt}\arabic{footnote}}
 \makeatletter \@addtoreset{footnote}{section} \makeatother

\section{\hspace*{-.85em}. Leptonic decay,
  \mbox{\boldmath$l^+l^-$}-annihilation} \label{App:llbar}
\setcounter{equation}{0}

Vector meson leptonic decay width and $l^+l^-$-annihilation cross section into
the final states $f \!=\! \pi^+\pi^-$ and $2\pi^+2\pi^-$ are determined by the same
$S$-matrix element:
\beq
S = \langle l^-(p,s)\, l^+(p',s') \left| S \right| V(q,\la) \rangle \;,
\end{equation}
where $p$, $p'$ and $q$ are the momenta, $s$, $s'$ and $\la$ the spins and
the helicity, respectively. With the $T$-matrix given through
\mbox{$S =: i(2\pi)^4 \de_4(p\!+\!p'\!-\!q)\; T$\;} we have
\beqa
T &=& -e\, \bar{u}_s(p)\ga^\mu v_{s'}(p')\;
  \frac{g_{\mu\nu}}{(p + p')^2 + i\varep}\;
  \langle0\left|J_{em}^{\nu}(0)\right|V(q,\lambda)\rangle \;, \\
&=& -e^2 \bar{u}_s(p)\ga^\mu v_{s'}(p')\;
  \frac{1}{s}\; f_V M_V \varep_{\mu}(q,\la) \;; \nn
\end{eqnarray}
in the last line we have introduced the total energy squared $s\!=\!(p\!+\!p')^2$
and the coupling $f_V$ of the vector meson to the electromagnetic current, which is
defined through
\beq \label{fV_coupling}
\langle0\left|J_{\rm em}^{\mu}(0)\right|V(q,\la)\rangle=ef_V M_V\varep^{\mu}(q,\la) \;.
\end{equation}

Averaging over incoming spins $s$ and $s'$ and summation over outgoing
helicities $\la$ gives
\beqa
\sum\nolimits' |T|^2 &=& -\frac{e^4}{3}\;
  \left(\frac{f_V M_V}{s}\right)^{\!\!2}\;
    \sum_{s,s'}\; {\rm tr}
    \left[ \ga^\mu u_s(p) \bar{u}_s(p) \ga_\mu v_{s'}(p') \bar{v}_{s'}(p') \right] \;, \\
&=& +\frac{4e^4}{3}\;
  \left(\frac{f_V M_V}{s}\right)^{\!\!2}\; s\cdot
    {\textstyle \left(1 + \frac{2m_l^2}{s}\right)} \;, \nn
\end{eqnarray}
where $m_l$ is the lepton mass.

The decay rate of the vector meson in its rest frame is
\beq
d\Ga = \frac{1}{2M_V}\; (2\pi)^4 \de_4(p + p' - q)\;
  \frac{d^3\vec{p}}{(2\pi)^3 2p_{0+}}
  \frac{d^3\vec{p}^{\,\prime}}{(2\pi)^3 2p^{\,\prime}_{0+}}\;
  \sum\nolimits' |T|^2 \;;
\end{equation}
phase space integration leads to
\beq \label{leptonic_width}
\Ga_{V\to l^+l^-} = \frac{4\pi\alpha^2}{3} \frac{f_V^2}{M_V}
  \cdot {\textstyle \left(1 + \frac{2m_l^2}{M_V^2}\right)
    \sqrt{1 - \frac{4m_l^2}{M_V^2}}} \;.
\end{equation}

The differential cross section to produce in $l^+l^-$-annihilation a real vector
meson $V$ is
\beq
d\si = \frac{1}{2w(s,m_l^2,m_l^2)}\; (2\pi)^4 \de_4(p + p' - q)\;
  \frac{d^3\vec{q}}{(2\pi)^3 2q_{0+}}\;
  \sum\nolimits' |T|^2 \;,
\end{equation}
where $w(x,y,z)$ is the K\"allen function; after phase space integration it
writes
\beq \label{annihilation}
\si_{l^+l^-\rightarrow V}
= \frac{4\pi e^4}{3} \left(\frac{f_V}{M_V}\right)^{\!\!2}\; \de(s - M_V^2)
  \cdot \frac{1 + \frac{2m_l^2}{M_V^2}}{\sqrt{1 - \frac{4m_l^2}{M_V^2}}} \;.
\end{equation}
For electrons the $m_l$-depending factors in
Eqs~(\ref{leptonic_width})~and~(\ref{annihilation}) can be neglected.

To calculate the $\rh$-channel mass spectra of photoproduction and
$e^+e^-$-annihilation into two and four charged pions we distribute the
$\rh$-, $\rh'$- and $\rh''$-mesons according to simple Breit-Wigner resonances:
\beq
T(M^2) := T \cdot \sqrt{\frac{M_V\Ga_V^{tot}}{\pi}}\;
  \frac{c_{V\!f}}{M^2 -M_V^2 + i M_V\Ga_V^{tot}} \;,
\end{equation}
where $T$ is the corresponding $T$-amplitude. The constraint for the
constants $c_{V\!f}$ is that the integrated cross section is not altered:
\beq \label{width_norm}
\int_{s_f}^\infty dM^2\, \left|T(M^2)\right|^2 = \left|T\right|^2 \;,
\end{equation}
where the threshold $s_f$ is either $(2m_\pi)^2$ or $(4m_\pi)^2$ for the
respective final state. The factor $|T|^2$ drops out and it follows
\beq \label{c_Vf}
c_{V\!f} = \left[ {\textstyle \frac{1}{2}
  + \frac{1}{\pi}\arctan\frac{M_V^2 - s_f}{M_V} } \right]^{-1\!/\!2} \;,
\end{equation}
which numerically implies values exceeding 1 up to seven percent.

We finally have
\beq \label{ann_mass}
\si_{e^+e^-\rightarrow f}(M^2) = \frac{4\pi e^4}{3}
  \left| \sum_{V\!=\rh,\rh',\rh''} \frac{f_V}{M_V}\;
  \sqrt{\frac{M_V\Ga_V^{tot}}{\pi}}\;
  \frac{c_{V\!f}}{M^2 -M_V^2 + i M_V\Ga_V^{tot}}\;
  \sqrt{B_{V\rightarrow f}}\; \right|^2
\end{equation}
(for the branching ratios, widths and masses cf. Table~\ref{Tabl:properties}).
For annihilation into $f\!=\!\pi^+\pi^-$ we parametrize in Eq.~(\ref{ann_mass}) the
$\rh$-width $\Ga_\rh^{tot}$ by the polynomial
{\small
\beq \label{Garh_polynomial}
\Ga_\rh^{tot}\;
  \left[ {\textstyle 1 + a_1 \cdot \left(\frac{M^2}{M_\rh^2} - 1\right)
                       + a_2 \cdot \left(\frac{M^2}{M_\rh^2} - 1\right)^2 }\right]
\end{equation}}
and adjust $a_1$, $a_2$ and $c_{\rh,\pi^+\pi^-}$ in order to reproduce the
experimental spectrum.

\section{\hspace*{-.85em}. Wave functions} \label{App:wave_fns}
\setcounter{equation}{0}

\paragraph{Construction.} In previous work, see App.~A in Ref.~\cite{G1}, we
explicitely constructed the photon wave function in the frame of light-cone
perturbation theory. For the probability amplitude for a photon with momentum
$q \!=\! \left( q^+, q^-\!=\!-Q^2\!/\!2q^+, {\bf q}={\bf 0} \right)$,
virtuality $Q$ and helicity $\la$ to fluctuate into a $q\bar q$-pair one has to
calculate the
expression\footnote{Throughout this appendix we abbreviate $\bar{z}\!=\!1\!-\!z$.}
\beq \label{ansatz_photon}
\tilde{\ps}^{h,\bar h}_{\ga(Q^2,\la)}(z,{\bf k}) \;=\;
\sqrt{N_c} \; e_f\de_{f\bar f} \;
  \frac{\sqrt{z\bar z}}{z\bar{z}Q^2 + m^2 + {\bf k}^2}\;
  \bar{u}(zq^+,{\bf k},h)\, \varep^\mu(q,\la)\ga_\mu\, v(\bar{z}q^+,-{\bf k},\bar{h}) \;,
\end{equation}
where the quark carries $zq^+$ longitudinal, $\bf k$ transverse momentum and
helicity $h$ (the antiquark accordingly $\bar{z}q^+$, $-\bf k$ and $\bar{h}$). 
With the polarization vectors
\mbox{$\varep(q,0) \!=\! \left( q^+\!/\!Q, Q\!/\!2q^+,{\bf 0} \right)$} and
\mbox{$\varep(q,\pm1) \!=\! -1\!/\!\sqrt{2}\, (0, 0, 1, \pm i)$} and the convention
of light-cone components \mbox{$q^\pm \!=\! (q^0 \!\pm\! q^3)\!/\!\sqrt{2}$},
$g_{+-} \!=\! 1$, a lengthy but straightforward evaluation of
Eq.~(\ref{ansatz_photon}) gives
\beqa \label{photon_pspace}
\tilde{\ps}^{h,\bar h}_{\ga(Q^2,\la)}(z,{\bf k}) =
  \sqrt{N_c} \; e_f\de_{f\bar f} \;
  &\hspace*{-2em}\bigg\{\hspace*{-2em}&
  -2z\bar{z}\, Q\, \delta_{h,-\bar h} \cdot \de^0_\la \\
&&+ \sqrt{2} \left[\pm k {\rm e}^{\pm i\varph_{\bf k}}\;
                \left( z \delta_{h+,\bar h-} - \bar{z} \delta_{h-,\bar h+} \right)
                + m\; \delta_{h\pm,\bar h\pm} \right] \cdot \de^\pm_\la\;
  \bigg\}\; \frac{1}{\varep^2 + k^2} \;, \nn
\end{eqnarray}
where $k\!=\!|{\bf k}|$ and \mbox{$\varep \!=\! \sqrt{z\bar{z} Q^2 \!+\! m^2}$},
cf. Eq.~(\ref{epsilon}).

For an arbitrary function $\tilde{f}({\bf k})$ we define the Fourier transform with
respect to the transverse momentum ${\bf k}$ through
\beq \label{FT}
\int \frac{d^2{\bf k}}{(2\pi)^2}\;
  {\rm e}^{i{\bf k}\cdot{\bf r}}\; \tilde{f}({\bf k})
  = f({\bf r}) \;.
\end{equation}
For an arbitrary function we have further:
\beq \label{FT_vector}
\int \frac{d^2{\bf k}}{(2\pi)^2}\; {\rm e}^{i{\bf k}\cdot{\bf r}}\;
  k {\rm e}^{\pm i\varph_{\bf k}}\; \tilde{f}({\bf k})
  = -i\, (\partial_1 \pm i\partial_2)\; f({\bf r}) \;.
\end{equation}
If $\tilde{f}$ does not depend on the direction of ${\bf k}$, the r.h.s. of
Eq.~(\ref{FT_vector}) can be written as
\beq
-i\, {\rm e}^{\pm i\varph_{\bf r}}\; \partial_r\; f(r) \;,
\end{equation}
where $r\!=\!|{\bf r}|$.

We thus have for the Fourier transform of Eq.~(\ref{photon_pspace}):
\beqa \label{photon_xspace1}
\hspace*{-.25em}
\ps^{h,\bar h}_{\ga(Q^2,\la)}(z,{\bf r}) &\hspace*{-.75em}=\hspace*{-.75em}&
  \sqrt{N_c} \; e_f\de_{f\bar f}\; \bigg\{
  -2z\bar{z}Q\, \delta_{h,-\bar h} \cdot \de^0_\la \\
&&\hspace*{6em} + \sqrt{2} \left[\pm i\, {\rm e}^{\pm i\varph_{\bf r}}\,
                \left( z \delta_{h+,\bar h-} - \bar{z} \delta_{h-,\bar h+} \right)\,
                  (-\partial_r)\;
                + m\; \delta_{h\pm,\bar h\pm} \right] \cdot \de^\pm_\la\; \bigg\} \nn \\
&&\hspace*{23em} \times \int \frac{d^2{\bf k}}{(2\pi)^2}\;
    {\rm e}^{i{\bf k}\cdot{\bf r}}\; \frac{1}{\varep^2 + k^2} \;. \nn  
\end{eqnarray}
With
\beq
\int \frac{d^2{\bf k}}{(2\pi)^2}\; {\rm e}^{i{\bf k}\cdot{\bf r}}\;
 \frac{1}{\varep^2 + k^2} = \frac{K_0(\varep r)}{2\pi}
\end{equation}
and $-\frac{d}{dz} K_0(z) \!=\! K_1(z)$ Eq.~(\ref{photon_xspace1}) becomes:
\beqa \label{photon_xspace2}
\ps^{h,\bar h}_{\ga(Q^2,\la)}(z,{\bf r}) &\hspace*{-.75em}=\hspace*{-.75em}&
  \sqrt{N_c} \; e_f\de_{f\bar f} \\
&& \times \bigg\{
  -2z\bar{z}Q\, \delta_{h,-\bar h}\, \frac{K_0(\varep r)}{2\pi} \cdot \de^0_\la \nn \\
&&\hspace*{2em} + \sqrt{2} \left[\pm i\varep\, {\rm e}^{\pm i\varph_{\bf r}}\;
                \left( z \delta_{h+,\bar h-} - \bar{z} \delta_{h-,\bar h+} \right)\,
                  \frac{K_1(\varep r)}{2\pi}\;
                + m\; \delta_{h\pm,\bar h\pm}\, \frac{K_0(\varep r)}{2\pi} \right]
                \cdot \de^\pm_\la\; \bigg\} \;, \nn  
\end{eqnarray}
which is Eqs~(\ref{photon1}) and (\ref{photon2}).

We model wave functions for the vector mesons according to the photon
wave function:
\beqa \label{vmeson_pspace}
\tilde{\ps}^{h,\bar h}_{V(\la)}(z,{\bf k}) \;=\;
  &\hspace*{-2em}\bigg\{\hspace*{-2em}&
  4z\bar{z}\, \om_{V,\la}\, \delta_{h,-\bar h} \cdot \de^0_\la \\
&&+ \left[\pm k {\rm e}^{\pm i\varph_{\bf k}}\;
                \left( z \delta_{h+,\bar h-} - \bar{z} \delta_{h-,\bar h+} \right)
                + m\; \delta_{h\pm,\bar h\pm} \right] \cdot \de^\pm_\la\;
  \bigg\}\; \tilde{\ps}_{V(\la)}(z,k) \;, \nn
\end{eqnarray}
where the energy denominator of the photon,
\mbox{$(z\bar{z}Q^2 \!+\! m^2 \!+\! {\bf k}^2)^{-1}$},
has been replaced by functions \mbox{$\tilde{\ps}_{V(\la)}(z,k)$} which also
do not depend on the direction of ${\bf k}$. We define for the $1S$-state
\beqa \label{scalar1S}
\tilde{\ps}_{1(\la)}(z,k)
&=& {\cal N}_{1,\lambda}\; \sqrt{z\bar{z}}\;
  {\rm e}^{ -\frac{1}{2}\, M^2\, \om_{1,\lambda}^{-2}\, (z-1/2)^2 } \cdot
  \frac{2\pi}{\om_{1,\lambda}^2}\, {\rm e}^{-\frac{1}{2} \om_{1,\lambda}^{-2} k^2} \;, \\
&=& h_{1, \lambda}(z) \cdot
  \frac{2\pi}{\om_{1,\lambda}^2}\, {\rm e}^{-\frac{1}{2} \om_{1,\lambda}^{-2} k^2} \nn
\end{eqnarray}
the harmonic oscillator parametrization by Wirbel and Stech, cf. Ref.~\cite{WS},
which is peaked at the nonrelativistic value $z \!=\! 1\!/\!2$.
For the $2S$-state we have to differentiate that it can be "radially" excited
either in longitudinal or, with two modes, in transverse direction.
We thus introduce the simplest polynomial which is symmetric under exchange of
$z\!\leftrightarrow\bar{z}$ in longitudinal direction and the transverse
dependence of the $2S$-harmonic oscillator:
\beqa \label{scalar2S}
\tilde{\ps}_{2(\la)}(z,k)
&=& {\cal N}_{2,\lambda}\; \sqrt{z\bar{z}}\;
  {\rm e}^{ -\frac{1}{2}\, M^2\, \om_{2,\lambda}^{-2}\, (z-1/2)^2 } \\
&&\hspace*{3.125em} \cdot
  \frac{2\pi}{\om_{2,\lambda}^2}\, {\rm e}^{-\frac{1}{2} \om_{2,\lambda}^{-2} k^2}\;
  \big\{ (z\bar{z} - A_\la)
    + \sqrt{2}\, (1 - \om_{2,\lambda}^{-2} k^2) \big\} \;, \nn \\
&=& h_{2, \la}(z) \cdot
  \frac{2\pi}{\om_{2,\lambda}^2}\, {\rm e}^{-\frac{1}{2} \om_{2,\lambda}^{-2} k^2}\;
  \big\{ (z\bar{z} - A_\la)
    + \sqrt{2}\, (1 - \om_{2,\lambda}^{-2} k^2) \big\} \nn \;. 
\end{eqnarray}
In Eqs~(\ref{scalar1S}) and (\ref{scalar2S}) we have used the definition of
$h_{V, \la}(z)$ from Eq.~(\ref{hg_defs}), factors in which
Eq.~(\ref{vmeson_pspace}) differs from Eq.~(\ref{photon_pspace}) are absorbed in the
normalization constants ${\cal N}_{V,\la}$.

Fourier transformation of Eq.~(\ref{vmeson_pspace}) gives
\beqa \label{vmeson_xspace1}
\ps^{h,\bar h}_{V(\la)}(z,{\bf r}) \;=\;
  &\hspace*{-2em}\bigg\{\hspace*{-2em}&
  4z\bar{z}\, \om_{V,\la}\, \delta_{h,-\bar h} \cdot \de^0_\la \\
&&+ \left[\pm i\, {\rm e}^{\pm i\varph_{\bf r}}\,
                \left( z \delta_{h+,\bar h-} - \bar{z} \delta_{h-,\bar h+} \right)\,
                  (-\partial_r)\;
                + m\; \delta_{h\pm,\bar h\pm} \right] \cdot \de^\pm_\la\; \bigg\}\;
                \ps_{V(\la)}(z,r) \;. \nn  
\end{eqnarray}
Using
\beq
\int \frac{d^2{\bf k}}{(2\pi)^2}\; {\rm e}^{i{\bf k}\cdot{\bf r}} \cdot
  \frac{2\pi}{\om_{1,\lambda}^2}\, {\rm e}^{-\frac{1}{2} \om_{1,\lambda}^{-2} k^2}
\;=\; {\rm e}^{-\frac{1}{2} \om_{1,\lambda}^2 r^2}
\end{equation}
and
\beq
\int \frac{d^2{\bf k}}{(2\pi)^2}\; {\rm e}^{i{\bf k}\cdot{\bf r}} \cdot
  \frac{2\pi}{\om_{1,\lambda}^2}\, {\rm e}^{-\frac{1}{2} \om_{1,\lambda}^{-2} k^2}\,
  (1 - \om_{1,\lambda}^{-2} k^2)
\;=\; {\rm e}^{-\frac{1}{2} \om_{1,\lambda}^2 r^2}\,
  (\om_{1,\lambda}^2 r^2 - 1)
\end{equation}
we obtain the representations given in Eqs~(\ref{wfnRhL}),~(\ref{wfnRhT}) and
Eqs~(\ref{wfn2SL}),~(\ref{wfn2ST}). 

\paragraph{Fixing of the parameters.} There are several parameters $\om_{V,\la}$,
${\cal N}_{V,\la}$ and $A_\la$ to be fixed. The constraints are as follows.

The first condition concerns the coupling to the electromagnetic current $f_V$,
see Eq.~(\ref{fV_coupling}), which is connected with the wave function at
the origin and determined by the vector meson $e^+e^-$-decay width through
\beq \label{eebar_width}
\Ga_{V\to e^+e^-} = \frac{4\pi\alpha^2}{3} \frac{f_V^2}{M_V} \;;
\end{equation}
cf. Eq.~(\ref{leptonic_width}) where the electron mass is neglected. With the
wave functions given this means for $\la\!=\!L,\;\pm1$:
\beqa \label{cond_fV}
f_{V,L} &=& \hat{e}_V \sqrt{N_c} \cdot 4\, \om_{V,L} \cdot
  \int \frac{dz d^2{\bf k}}{16\pi^3}\; 4z\bar{z} \cdot \ps_{V(L)}(z,k) \;, \\
f_{V,T} &=& \hat{e}_V \sqrt{N_c} \cdot\, \frac{4\sqrt{2}}{M_V} \cdot\,
  \int \frac{dz d^2{\bf k}}{16\pi^3}\;
  \left\{ (z^2 + \bar{z}^2) k^2 + m^2 \right\}\;
  \frac{1}{4z\bar{z}} \cdot \tilde{\ps}_{V(\la=\pm1)}(z,k) \;, \nn
\end{eqnarray}
where $\hat{e}_V$ denotes the effective quark charge in the vector meson $V$
in units of the electromagnetic charge, i.e. $\hat{e}_V \!=\! 1\!/\!\sqrt{2}$
for the $\rh$-mesons. Note, that in the transverse case the mass of the
corresponding state enters explicitely; the numerical value of $f_{2T}$ given
in Table~\ref{Tabl:wf_cnsts} is based on $M_{2S}\!=\!1.6\GeV$.

The second condition is the normalization of the wave functions according to
\beq
\langle V(q',\la^{\prime})|V(q,\la)\rangle = (2\pi)^3 2q^+ \de(q^+-q^{\prime+})
  \de_2({\bf q}-{\bf q^{\prime}}) \de_{\la\la^{\prime}} \;,
\end{equation}
i.e.
\beq
1 = \int \frac{dz d^2{\bf k}}{16\pi^3}\;
  \sum_{h,\bar{h}} \left| \tilde{\ps}^{h,\bar h}_{V(\la)}(z,{\bf k}) \right|^2 \;;
\end{equation}
or explicitly for $\la\!=\!L,\;\pm1$:
\beqa \label{cond_norm}
1 &=& 2\, \om_{V,L}^2\cdot \int \frac{dz d^2{\bf k}}{16\pi^3}\;
  (4z\bar{z})^2\; \left| \tilde{\ps}_{V(L)}(z,k) \right|^2 \;, \\
1 &=& 2\cdot \int \frac{dz d^2{\bf k}}{16\pi^3}\;
  \left\{ (z^2 + \bar{z}^2) k^2 + m^2 \right\}\;
  \left| \tilde{\ps}_{V(\la =\pm1)}(z,k) \right|^2 \;. \nn
\end{eqnarray}

We first turn to the $1S$-state. Taken as input the experimentally well-determined
quantities $f_\rh$ and $M_\rh$, cf. Table~\ref{Tabl:properties}, we have for each helicity
the set of implicit equations~(\ref{cond_fV})~and~(\ref{cond_norm}) to determine
$\om_{1,\la}$ and ${\cal N}_{1,\la}$.

For the $2S$-state we exploit both the conditions of normalization, Eq.~(\ref{cond_norm}),
and orthogonality to the $1S$-state:
\beqa \label{cond_ortogonal}
0 &=& \int \frac{dz d^2{\bf k}}{16\pi^3}\; (4z\bar{z})^2\;
  \tilde{\ps}_{1(L)}^{\dagger}(z,k) \tilde{\ps}_{2(L)}(z,k) \\
0 &=& \int \frac{dz d^2{\bf k}}{16\pi^3}\;
  \left\{ (z^2 + \bar{z}^2) k^2 + m^2 \right\}\;
  \tilde{\ps}_{1(\la=\pm1)}^{\dagger}(z,k) \tilde{\ps}_{2(\la=\pm1)}(z,k) \;. \nn
\end{eqnarray}
If one would set $\om_{2,\la} \!=\! \om_{1,\la}$ and determine ${\cal N}_{2,\la}$
and $A_\la$, the results for $f_{2,\la}$ from Eq.~(\ref{cond_fV}) would not coincide for
longitudinal and transverse polarization.
However, it is possible to obtain agreement $f_{2L} \!=\! f_{2T}$, if one allows for
slight deviation of $\om_{2L}$ away from $\om_{1L}$ and $\om_{2T}$ away from $\om_{1T}$.

We refer to Table~\ref{Tabl:wf_cnsts}, where we list the numerical
values of the so-determined parameters.

\section{\hspace*{-.85em}. Mixing angle, \mbox{\boldmath$f_{2S}$}-coupling} \label{App:mixing}
\setcounter{equation}{0}

We make some remarks on Table~\ref{Tabl:properties} and the derivation
of the mixing angle $\th$ from the branching ratios $X_1$, $X_2$ and $X_3$. 

With our simple ansatz for the states $\rh(1450)$ and $\rh(1700)$ as mixtures
of a quark-antiquark $2S$-state and an inert rest, cf. Eq.~(\ref{mixing}), we have
\beqa
f_{\rh'}  &=& \;\;\; \cos\th \; f_{2S} \;, \\
f_{\rh''} &=&       -\sin\th \; f_{2S} \;. \nn
\end{eqnarray}
For the leptonic decay widths, see Eq.~(\ref{leptonic_width}), this means
\beqa \label{lept_widths1}
\Ga_{\rh' \rightarrow e^+e^-}
  &=& \frac{4\pi\al^2}{3}\, f_{2S}^2\, \frac{\cos^2\th}{M_{\rh'}} \;, \\
\Ga_{\rh'' \rightarrow e^+e^-}
  &=& \frac{4\pi\al^2}{3}\, f_{2S}^2\, \frac{\sin^2\th}{M_{\rh''}} \;. \nn
\end{eqnarray}
On the other side we have from Eqs~(\ref{branchings})
\beqa \label{lept_widths2}
\Ga_{\rh' \rightarrow e^+e^-}
  &=& \Ga_{\rh'}^{tot} \cdot \left.\frac{X_1(1\!+\!X_2)}{X_3}\right|_{\rh'} \;, \\
\Ga_{\rh'' \rightarrow e^+e^-}
  &=& \Ga_{\rh''}^{tot} \cdot \left.\frac{X_1(1\!+\!X_2)}{X_3}\right|_{\rh''} \;. \nn
\end{eqnarray}
Equating (\ref{lept_widths1}) and (\ref{lept_widths2}) we find
\beq
\tan^2\th
= \left. M_{\rh'} \Ga_{\rh'}^{tot}
          \cdot \left.\frac{X_1(1\!+\!X_2)}{X_3}\right|_{\rh'}
 \right/ M_{\rh''} \Ga_{\rh''}^{tot}
          \cdot \left.\frac{X_1(1\!+\!X_2)}{X_3}\right|_{\rh''}  \;.
\end{equation}
and
\beq
f_{2S}^2 = \frac{3}{4\pi\al^2}\,
  \left( M_{\rh'} \Ga_{\rh'}^{tot}
          \cdot \left.\frac{X_1(1\!+\!X_2)}{X_3}\right|_{\rh'}
       + M_{\rh''} \Ga_{\rh''}^{tot}
          \cdot \left.\frac{X_1(1\!+\!X_2)}{X_3}\right|_{\rh''}
  \right) \;,
\end{equation}
i.e. numerically
\beqa
\th &=& \;\; 41.2^\circ \;, \\
f_{2S} &=& -0.178 \GeV \;, \nn
\end{eqnarray}
where $f_{2S}$ is negative and the mixing angle is chosen in the first quadrant
to have the interference pattern in Fig.~\ref{Fig:eebar2pis}.

The coupling of the $2S$-state to the electromagnetic current apparently
differs from the value in Table~\ref{Tabl:wf_cnsts} which comes from our model
$2S$-wave functions, cf. Eqs~(\ref{wfn2SL}) and (\ref{wfn2ST}), which we require
to be normalized and orthogonal on the $1S$-states.
With regard to the accuracy of the numerical values of $X_1$, $X_2$ and $X_3$,
cf. Table~\ref{Tabl:properties} and Ref.~\cite{DM}, we feel legitimized to base
our calculation on the mixing angle derived above and the coupling in
Table~\ref{Tabl:wf_cnsts}, instead of adjusting the wave function parameters
in order to obtain global agreement.

\markright{References}


\begin{thebibliography}{99}
\bibitem{Aston1}   D.~Aston et al., Phys.Lett. {\bf B92} (1980) 215.
\bibitem{Aston2}   D.~Aston et al., Nucl.Phys. {\bf B189} (1981) 15 and {\bf B209} (1982) 56.
\bibitem{Q}        A.~Quenzer et al., Phys.Lett. {\bf B76} (1978) 512.
\bibitem{Ba}       L.M.~Barkov et al., Nucl.Phys. {\bf B256} (1985) 365.
\bibitem{Bi}       D.~Bisello et al., Orsay preprint, LAL: 85-15 (1985).
\bibitem{DM}       A.~Donnachie and H.~Mirzaie, Z.Phys. {\bf C33} (1987) 407.
\bibitem{CD}       A.B.~Clegg and A.~Donnachie, Z.Phys. {\bf C62} (1994) 455.
\bibitem{G1}       H.G.~Dosch, T.~Gousset, G.~Kulzinger and H.J.~Pirner,
                     Phys.Rev. {\bf D55} (1997) 2602.
\bibitem{G2}       H.G.~Dosch, T.~Gousset, H.J.~Pirner, Phys.Rev. {\bf D57} (1998) 1666.
\bibitem{JW}       J. Berges, D.~Jungnickel and C.~Wetterich, hep-ph/9705474.
\bibitem{SchP}     B.J.~Sch\"afer and H.J.~Pirner, hep-ph/9712413.
\bibitem{Close}    F.E.~Close and P.R.~Page, Nucl.Phys. {\bf B443} (1995) 233
                     and Phys.Rev. {\bf D56} (1997) 1584.
\bibitem{Be}       L.~Bergstr{\o}m, H.~Snellman and G.~Tengstrand, Phys.Lett. {\bf B80} (1979) 242.
\bibitem{Simula}   F.~Cardelli, I.L.~Grach, I.M.~Narodetskii, G.~Salme and S.~Simula,
                     Phys.Lett. {\bf B349} (1995)~393.
\bibitem{MoShSi}   V.L.~Morgunov, V.I.~Shevchenko and Yu.A.~Simonov, hep-ph/9704282.
\bibitem{WS}       M. Wirbel, B. Stech and M. Bauer, Z.Phys. {\bf C29} (1985) 637.
\bibitem{HaZh}     I. Halperin and A. Zhitnitsky, Phys.Rev. {\bf D56} (1997) 184.
\bibitem{BaBr}     P. Ball and V.H. Braun, Phys.Rev. {\bf D54} (1996) 2182.
\bibitem{Di}       M. Diehl, hep-ph/9707441.
\bibitem{DFK}      H.G.~Dosch, E.~Ferreira and A.~Kr\"amer, Phys.Rev. {\bf D50} (1994) 1992.
\bibitem{KZ}       B.Z.~Kopeliovich and B.G.~Zakharov, Phys.Rev. {\bf D44} (1991) 3466.
\bibitem{NNN}      J.~Nemcik, N.N.~Nikolaev, E.~Predazzi, and B.G.~Zahkarov,
                     Phys.Lett. {\bf B374} (1996) 199, Z.Phys. {\bf C75} (1997) 71 and J.~Nemcik, 
                     N.N.~Nikolaev, E.~Predazzi, B.G.~Zahkarov and V.R.~Zoller, hep-ph/9712469. 
\bibitem{NMC}      NMC-collaboration, M.~Arneodo et al., Nucl.Phys. {\bf B429} (1994) 503.
\bibitem{E665_1}   E665-collaboration, M.R.~Adams et al., MPI-PHE-97-03 Feb 1997,
                     submitted to Z.Phys.~{\bf C}.
\bibitem{DoLa}     A.~Donnachie and P.V.~Landshoff, Phys.Lett. {\bf B296} (1992) 227.
\bibitem{Adams}    M.R.~Adams et al., Phys.Rev.Lett. {\bf 74} (1994) 1525.
\bibitem{E687}     P.~Lebrun, E687-collaboration, Contribution to the 7th International Conference
                     on Hadron Spectroscopy, Hadron 97, Aug 25-30, 1997 BNL, Upton NY, to appear
                     in the proceedings.
\bibitem{H1}       H1-collaboration, paper pa01-088, submitted to ICHEP 96, Warsaw 1996 (Poland).
\bibitem{Atkinson} M.~Atkinson et al., Z.Phys. {\bf C30} (1986) 531.
\bibitem{PDG}      Particle Data Group: Phys.Rev. {\bf D54} (1996) 1.
\bibitem{E665_2}   E665-collaboration, C.~Salgado, CEBAF-workshop on colour transparency (1995).
\end{thebibliography}
\end{document}